\documentclass[12pt,preprint]{aastex}
\usepackage{ulem}
\usepackage{epsfig}
\usepackage{epstopdf}
\usepackage{natbib}
\usepackage{verbatim}
\usepackage{amsmath}
\usepackage{graphicx}
\usepackage{color}

\newcommand{\ha}{H$\alpha$ }
\newcommand{\Spitzer}{{\sl Spitzer} }
\DeclareMathOperator{\TiOindex}{TiO-7140}
\slugcomment{}

\shorttitle{Searching for SBs within TDs}
\shortauthors{Kohn et al.\ }

\begin{document}


\title{SEARCHING FOR SPECTROSCOPIC BINARIES WITHIN TRANSITION DISK OBJECTS\footnote{This paper is based on data gathered with the 6.5 m Clay Telescope located at Las Campanas Observatory, Chile.}}

\author{Saul~A.~Kohn}
\affil{Lowell Observatory, 1400 W. Mars Hill Road, Flagstaff, AZ 86001, USA}
\affil{Department of Physics and Astronomy, University of Pennsylvania, Philadelphia, PA, 19104, USA}
\email{saulkohn@sas.upenn.edu}

\author{Evgenya~L.~Shkolnik}
\affil{School of Earth and Space Exploration, Arizona State University, Tempe, AZ 85287, USA}
\affil{Lowell Observatory, 1400 W. Mars Hill Road, Flagstaff, AZ 86001, USA}
\email{shkolnik@asu.edu}

\author{Alycia~J.~Weinberger}
\affil{Department of Terrestrial Magnetism, Carnegie Institution of Washington, 5241 Broad Branch Road, NW, Washington, DC 20015, USA}
\email{weinberger@dtm.ciw.edu}

\author{Joleen~K.~Carlberg}
\affil{NASA/GSFC Code 667, Greenbelt, MD 20771, USA}
\email{joleen.k.carlberg@nasa.gov}

\and

\author{Joe~Llama}
\affil{Lowell Observatory, 1400 W. Mars Hill Road, Flagstaff, AZ 86001, USA}
\affil{SUPA, School of Physics \& Astronomy, North Haugh. St Andrews. Fife. KY16 9SS, UK}
\email{joe.llama@st-andrews.ac.uk}

\begin{abstract}
Transition disks (TDs) are intermediate stage circumstellar disks characterized by an inner gap within the disk structure. 
To test whether these gaps may have been formed by closely orbiting, previously undetected stellar companions, we collected high-resolution optical spectra of 31 TD objects to search for spectroscopic binaries (SBs). 
Twenty-four of these objects are in Ophiuchus and seven are within the Coronet, Corona Australis, and Chameleon I star-forming regions. 
We measured radial velocities for multiple epochs, obtaining a median precision of 400 ms$^{-1}$.
We identified double-lined SB SSTc2d J163154.7-250324 in Ophiuchus, which we determined to be composed of a K7($\pm$0.5) and a K9($\pm$0.5) star, with orbital limits of $a<$0.6 AU and $P<$150 days.
This results in an SB fraction of 0.04$^{+0.12}_{-0.03}$ in Ophiuchus, which is consistent with other spectroscopic surveys of non-TD objects in the region. This similarity suggests that TDs are not preferentially sculpted by the presence of close binaries and that planet formation around close binaries may take place over similar timescales to that around single stars.
\end{abstract}


\section{Introduction}\label{intro}

Transition disks (TDs) are an intermediate stage of circumstellar disk evolution among pre-main sequence (PMS) stars. They are characterized by a lack of near-infrared excesses caused by inner optically thin holes or `gaps' opening within the disk structure, which typically have radii between one and tens of AU, and large far-infrared excesses from the disk material at larger separations \citep[e.g.,][]{Hughes.07, Hughes.09, Andrews.11,Zhu.11}.

Planet formation and dust aggregation are thought to cause the gaps observed in TDs, with inner holes greater than 15 AU possibly due to the formation of multiple planets \citep{Dodson-Robinson.11, Zhu.11}.
Other gap-formation mechanisms may involve ultraviolet (UV) and X-ray photons irradiating the disk and accretion from the disk onto the host star \citep[e.g.][]{Gorti.09, Espaillat.12}. Recent observations by \cite{Follette.13} support the results of simulations from, e.g. \cite{Rice.06} and \cite{Zhu.12} that dust dynamics may play a role in gap formation. The distribution of large and small grains in the inner regions of disks can be modified by gas pressure gradients, perhaps also generated by planets, that may trap or filter large grains as well as allow grains to grow efficiently \citep[e.g.][]{Zhu.11, Pinilla.12}.

A closely orbiting stellar companion may also produce an inner gap in a circumstellar disk \citep{Lubow94, D.Alessio.05}, as illustrated by the  CoKu Tauri/4 system. 
This system was observed to have a 10 AU wide gap \citep{Forrest.04}. The initial interpretation for such a wide gap was that a 10 $M_{\rm Jup}$ planet orbiting the central star cleared out the material \citep{Quillen.04}.
However, it was afterward shown that CoKu Tauri/4 was in fact a close visual binary (VB) system with a separation of $\sim$8 AU \citep{Kraus08}, and that the secondary star likely carved out the gap. Similarly, \cite{Biller.12} also detected a VB within the TD object HD 142527. With a projected separation of (12.8$\pm$1.5\,AU), the companion is likely to have played a role in shaping the system's complex disk structure, and it may have been responsible for the disk's cleared inner ring.

A close binary system need not preclude circumbinary planet formation. Using photometric data from \textit{Kepler}, \cite{Doyle.11} announced the first discovery of a circumbinary planet. 
Several other circumbinary planets have since been discovered \citep[e.g.][]{Orosz.12, Qian2.12, Qian.12, Schwamb.13}. 
These examples illustrate the need to further study multiple-star systems that host circumbinary material, and to test the effects of companion stars on disk dissipation and planet formation. Young star-forming regions are good test beds for such a study.

The nearby (140 pc) Ophiuchus star-forming region (Oph SFR) contains a population of young stars of an average age $\sim$2.2 Myr \citep{Wilking.05,M.-LombardiAlves08,Erickson.11}.
More than 50\% of the young stellar objects (YSOs) in Oph show evidence of optically thick circumstellar disks at mid-IR wavelengths \citep{Bontemps.01}, of which 9\% are identified as TDs \citep{Evans.03,Cieza.10}. This makes it well-suited for studies of disk dissipation.

Previous surveys of multiplicity in Oph and other SFRs have concentrated on visual binarity across all of the known active star-forming regions \citep[e.g.\ Table 6 of][]{Lafreniere.08}. With the exception of interferometric methods that are sensitive to separations of order 0.1--1 AU \citep{Pott.10}, VB searches are typically limited to separations $>$4 AU \citep{Ghez.93, Kraus.12}. \cite{Cieza.10} (hereafter C10) used IR and optical data to identify the TDs in Oph.
C10 also sub-classify the TDs in their sample into five catagories: grain growth-dominated disks (which are accreting and have a negative slope of IR excess: 44\% of their sample), giant planet-forming disks (accreting with a positive slope of IR excess: 13\%), photoevaporating disks (non-accreting with high disk luminosity: 17\%), disks already in the debris disk stage (non-accreting with low disk luminosity: 13\%) and circumbinary disks (13\%).
C10 did not observe an increase in detected companions within the separations to which their survey was most sensitive to (8--20 AU) compared with other separations, leading them to suggest that stellar companions at these separations are not responsible for a large fraction of the TD population. 

We present a complementary radial velocity (RV) survey to C10, to search for stellar companions closer in to the primary star. Our aim is to test whether the spectroscopic binary (SB) fraction is similar or enhanced in the TD population relative to stars without such disks.
The structure of this paper is as follows. 
The selection of the TD systems, observations and data reduction are described in Section~\ref{sec:sampobs}. 
RV, rotational velocity, and spectral type (SpT) measurements are presented in Section~\ref{res}. 
We report the discovery of a double-lined spectroscopic binary in Section~\ref{results}, and evaluate possible biases and the completeness of our survey in Section~\ref{complete}.
We derive a multiplicity fraction for Oph TDs, compare with previous multiplicity surveys of Oph, and discuss what multiplicity among TD objects might mean for theories of disk dispersal in Section~\ref{disc}. We summarize and conclude our study in Section~\ref{conc}.
While not the main objective of this paper, we provide values of lithium I (\ion{Li}{1}) and H$\alpha$ equivalent widths (EWs), and accretion rates for our sample, and a discussion on the unusual nature of Oph star SR21A in the Appendix.

\section{Sample and Observations}
 \label{sec:sampobs}
To survey the innermost regions of TD objects for stellar companions, we collected multi-epoch, high-resolution optical spectroscopy of 24 stars in Oph identified as having TDs \citep[C10;][]{AndrewsWilliams.07, Geers.07, Furlane.09}, two in Corona Australis (CrA; $<$3 Myr; \citealt{Forbrich08, S-A.08, Hughes.10}), two in Coronet (Cor, embedded within CrA; $<$1 Myr; \citealt{Knacke.73, S-A.08, Hughes.10}), and three in Chameleon I (Cha; 1--3 Myr; \citealt{Furlane.09, Kim.09}).

C10 identified Oph TDs from their colors as measured by the \Spitzer \textit{Cores 2 Disks} legacy program \citep{Evans.03}, and 20 of our 24 targets come from that work. A \Spitzer color (where $[x]$ indicates the magnitude in the $x\,\mu$m band) of $[3.6]-[4.5] < 0.25$ indicates the flux deficit characteristic of the near-infrared/inner opacity holes of a TD, while a color of $[3.6]-[24] > 1.5$ ensures that all of the targets have significant mid-IR excesses at 24$\,\mu$m. We included an additional four Oph targets and seven from other regions, all of which were identified by other authors using the same criteria as C10 \citep{AndrewsWilliams.07, Geers.07, Furlane.09}. 

\begin{figure}[htp]
\centering
\includegraphics[scale=0.7]{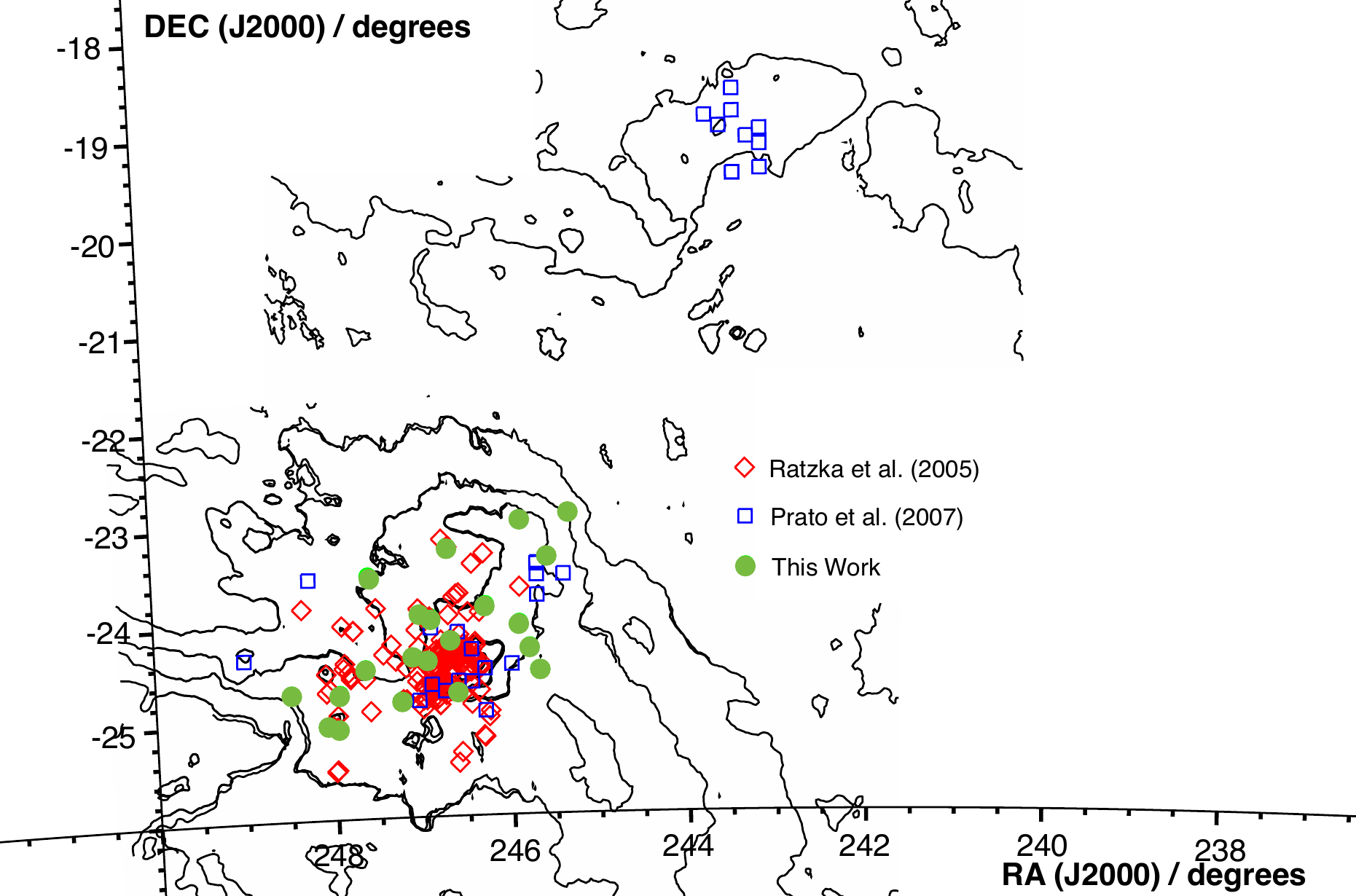}
\caption{R.A. and Dec.~of our targets within Ophiuchus, overlaid with other multiplicity surveys.
Contours are from the \textit{Infrared Astronomical Satellite} (IRAS) \textit{Sky Survey Atlas} at 100$\mu$m \citep{Neugebauer.84, Beichman.88}.}
\label{fig_Oph_map}
\end{figure}

\begin{figure}
\centering
\includegraphics[scale=0.5]{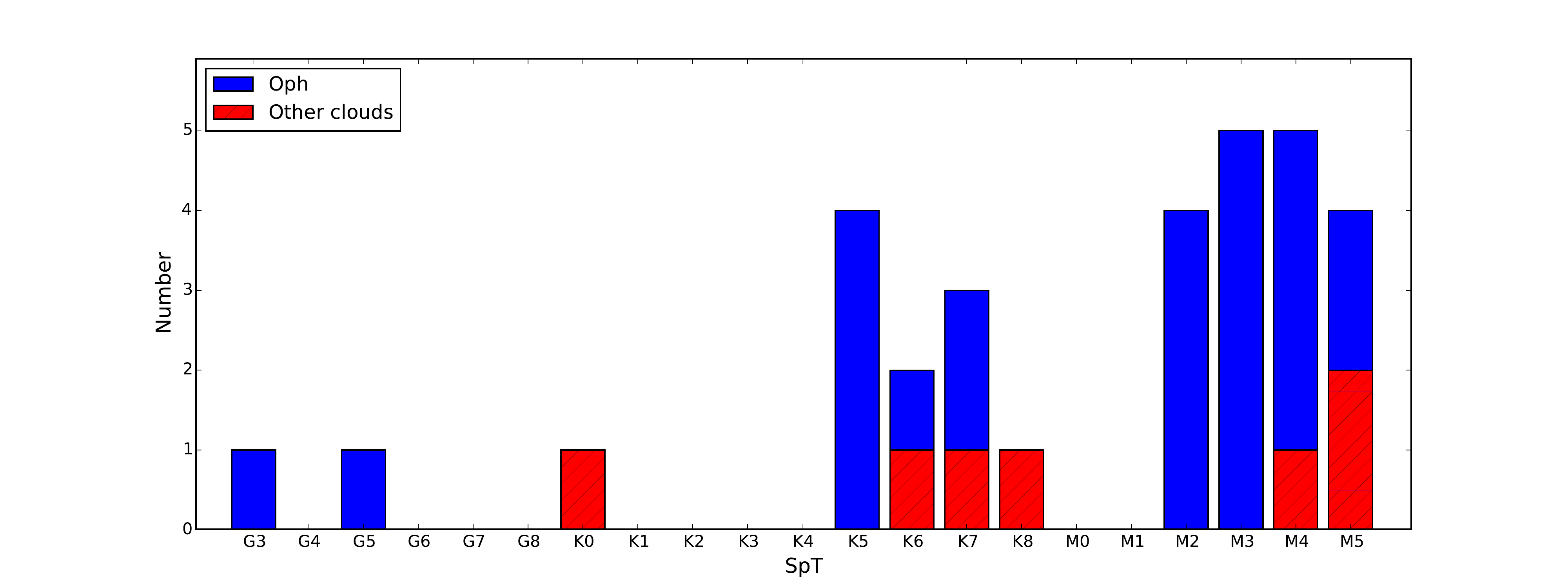}
\caption{The spectral type distribution of the TD sample. All of the targets have ages $\leq$10\,Myr.}
\label{fig_SpTs}
\end{figure}

The Oph targets in this study are shown in Figure~\ref{fig_Oph_map} in relation to other surveys of YSOs in the region. Table~\ref{table_td} lists each target in our sample, its stellar association, coordinates, SpT, and \textit{V}, \textit{R} and \textit{J} magnitudes. Targets that have known wide stellar companions are also noted.
The SpT distribution of our sample is shown in Figure~\ref{fig_SpTs} and is dominated by late-K to mid-M stars with a median SpT of M2. 

\clearpage
\begin{deluxetable}{llccclllcll} 
\tabletypesize{\scriptsize}
\rotate
\tablecaption{Transition Disk Targets \label{table_td}}
\tablewidth{0pt}
\tablehead{
\colhead{Name}		&	\colhead{Other}	&	\colhead{Association}	&	\colhead{$\alpha$ (J2000)} 			&	\colhead{$\delta$ (J2000)} 			&	\colhead{SpT} &	\colhead{\textit{V}}	&	\colhead{\textit{R}} &	\colhead{\textit{J} (2MASS)}	&	\colhead{Known Visual Multiplicity\tablenotemark{a}}	&	\colhead{Source\tablenotemark{b}}\\	
\colhead{}		&	\colhead{Identifier}	&	\colhead{}	&	\colhead{{h}\phn{m}\phn{s}}			&	\colhead{\phn{\arcdeg}~\phn{\arcmin}~\phn{\arcsec}}			&	\colhead{}	&	\colhead{mag} &	\colhead{mag}&	\colhead{mag}	&	\colhead{(Separation in \arcsec)}	&\colhead{}}	
\startdata																						
SZ Cha          	&	               	&	       Cha     	&	       10      58      16.8    	&	       -77     17      17.1    	&	       K0      	&	12.68	&		&	9.25	&	       Binary  ($\sim$5)	&	       (1),    (2) , (3), (4)\\
T25             	&	       Sz 18   	&	       Cha     	&	       11 07   19.1    	&	       -76     03      04.8    	&	       M2.5      	&	15.35	&	13.7	&	10.96	&		&	       (1), (5)\\   
T35             	&	       Sz 27   	&	       Cha     	&	       11      08      39.0    	&	       -77     16      04.2    	&	       K8      	&		&	15.82	&	11.17	&		&	       (1), (5)\\   
RX J1852.3-3700         	&	               	&	       CrA     	&	       18      52      17.3    	&	       -37     00      12.0    	&	       K7      	&	12.19	&		&	9.77	&		&	(7),(8)\\          
CrA-4111                	&	               	&	       CrA     	&	       19      01      20.8    	&	       -37     03      03.0    	&	       M4.5    	&		&		&	13.23	&		&	(9)\\          
G-49            	&	CrA 468	&	       Cor     	&	       19      01      49.4    	&	       -37     00      28.0    	&	       M4      	&		&	16.6	&	12.5	&		&	(9), (10)\\          
G-102           	&	CrA 133	&	       Cor     	&	       19      01      25.6    	&	       -37     04      53.0    	&	       M5      	&		&	15.3	&	12.36	&		&	(9), (5)\\          
\\\hline\\                                                                                                                                                                              																				
SSTc2d  J162118.5-225458        	&	               	&	       Oph     	&	       16      21      18.5    	&	       -22     54      58.0    	&	       M2      	&		&		&	11.45	&		&	(11)\\              
SSTc2d  J162218.5-232148        	&	               	&	       Oph     	&	       16      22      18.5    	&	       -23     21      48.0    	&	       K5      	&	12.67	&		&	9.52	&		&	(11), (12)\\          
SSTc2d  J162245.4-243124        	&	               	&	       Oph     	&	       16      22      45.4    	&	       -24     31      24.0    	&	       M3      	&		&	14.2	&	10.38	&		&	(11), (5)\\          
SSTc2d J162309.2-241705 	&	               	&	       Oph     	&	       16      23      09.3    	&	       -24     17      03.0    	&	       M?      	&	12.75	&	14.2	&	10.32	&		&	(1), (5)\\          
SSTc2d  J162332.8-225847        	&	               	&	       Oph     	&	       16      23      32.8    	&	       -22     58      47.0    	&	       M5      	&		&	15.7	&	11.49	&		&	(11), (5)\\          
SSTc2d  J162336.1-240221        	&	       	&	       Oph     	&	       16      23      36.1    	&	       -24     2       21.0    	&	       M5      	&	14.68	&		&	11.53	&		&	(11), (13)\\          
SSTc2d  J162506.9-235050        	&	               	&	       Oph     	&	       16      25      06.9    	&	       -23     50      50.0    	&	       M3      	&		&		&	11.05	&		&	(11)\\          
SSTc2d  J162623.7-244314        	&	       DoAr 25 	&	       Oph     	&	       16      26      23.7    	&	       -24     43      14.0    	&	       K5      	&		&	12.65	&	9.39	&		&	(11), (14)\\          
SSTc2d  J162646.4-241160        	&	               	&	       Oph     	&	       16      26      46.4    	&	       -24     11      60.0    	&	       G5      	&		&	13.9	&	9.68	&	       Binary  (0.58)	&	(11),   (14), (15)\\   
DoAr28          	&	       Haro 1-8        	&	       Oph     	&	       16      26      47.4    	&	       -23     14      52.2    	&	       K?      	&	13.84	&	12.1	&	9.89	&		&	(16), (17), (5)    \\      
SR21A           	&	               	&	       Oph     	&	       16      27      10.3    	&	       -24     19      12.7    	&	       G3      	&	14.1	&		&	8.75	&	       Binary  (896)	&	(18), (19), (20)\\   
SSTc2d  J162738.3-235732        	&	        DoAr 32        	&	       Oph     	&	       16      27      38.3    	&	       -23     57      32.0    	&	       K5      	&		&	13.66	&	9.91	&	       Triple(?)\tablenotemark{c}	&	(11), (14), (15), (21)\\
SSTc2d  J162739.0-235818        	&	       DoAr 33 	&	       Oph     	&	       16      27      39.0    	&	       -23     58      18.0    	&	       K6      	&		&	13.24	&	9.9	&	       Triple(?)\tablenotemark{c}	&	(11), (14), (15), (21)\\
SSTc2d  J162740.3-242204        	&	       DoAr 34 	&	       Oph     	&	       16      27      40.3    	&	       -24     22      04.0    	&	       K5      	&	11.5	&	11.1	&	8.44	&		&	(11), (22)\\
SSTc2d  J162802.6-235504        	&	               	&	       Oph     	&	       16      28      02.6    	&	       -23     55      04.0    	&	       M3      	&		&		&	11.76	&		&	(11)\\          
SSTc2d  J162821.5-242155        	&	               	&	       Oph     	&	       16      28      21.5    	&	       -24     21      55.0    	&	       M3      	&		&	16.97	&	12.08	&		&	(11), (14)\\          
SSTc2d  J162854.1-244744        	&	               	&	       Oph     	&	       16      28      54.1    	&	       -24     47      44.0    	&	       M2      	&		&	15.24	&	10.68	&		&	(11), (14)\\          
SSTc2d  J163020.0-233108        	&	               	&	       Oph     	&	       16      30      20.0    	&	       -23     31      08.0    	&	       M4      	&		&		&	11.32	&		&	(11)\\          
SSTc2d  J163033.9-242806        	&	               	&	       Oph     	&	       16      30      33.9    	&	       -24     28      06.0    	&	       M4      	&		&		&	11.63	&		&	(11)\\          
SSTc2d  J163145.4-244307        	&	               	&	       Oph     	&	       16      31      45.4    	&	       -24     43      07.0    	&	       M4      	&		&		&	11.8	&		&	(11)\\          
SSTc2d  J163154.4-250349        	&	               	&	       Oph     	&	       16      31      54.4    	&	       -25     03      49.0    	&	       M4      	&		&		&	11.78	&		&	(11)\\          
SSTc2d  J163154.7-250324        	&	    H$\alpha$74           	&	       Oph     	&	       16      31      54.7    	&	       -25     03      24.0    	&	       K7      	&	12.64	&	13.3	&	10.14	&		&	(11), (5)\\          
SSTc2d  J163205.5-250236        	&	               	&	       Oph     	&	       16      32      05.5    	&	       -25     02      36.0    	&	       M2      	&		&	15.5	&	11.65	&		&	(11), (5)\\          
SSTc2d  J163355.6-244205        	&	               	&	       Oph     	&	       16      33      55.6    	&	       -24     42      05.0    	&	       K7      	&		&	14.1	&	10.46	&		&	(11), (5)\\                       	
\enddata
\tablenotetext{a}{Binary companions all lie outside of the TD.}
\tablenotetext{b}{References: 
(1) \cite{Furlane.09},
(2) \cite{Kim.09},
(3) \cite{Ducati.02},
(4) \cite{Vogt.12},
(5) \cite{Cutri.03},
(6) \cite{Lopez.13},
(7) \cite{Hughes.10},
(8) \cite{Kiraga.12},
(9) \cite{S-A.06},
(10) \cite{Monet.03},
(11) \cite{Cieza.10},
(12) \cite{Vrba.93},
(13) \cite{Allers.06},
(14) \cite{Wilking.05},
(15)	\cite{Ratzka.05},
(16)	\cite{AndrewsWilliams.07},
(17) \cite{Samus.03},
(18)	\cite{Geers.07},
(19) \cite{Percheron02},
(20) \cite{Prato.03},
(21)	\cite{Barsony.05}. The SpTs are from the first reference listed, followed by the references for \textit{V} and \textit{R} magnitudes, and then visual multiplicity, if applicable. All \textit{J} magnitudes are from the 2MASS catalog \citep{Skrutskie.06}.}
\tablenotetext{c}{SSTc2d J162738.3-235732 and SSTc2d J162739.0-235818 are both possible members of the same triple system about primary ROXs 30A \protect\citep[16h\,27m\,37.0s -23\arcdeg 59\arcmin 32\arcsec ;][]{Ratzka.05}.} 
\end{deluxetable}
\clearpage

\subsection{Observations \label{obs}}

We acquired high-resolution optical spectra using the Magellan Inamori Kyocera Echelle (MIKE; \citealt{Bernstein.03}) spectrograph at the Magellan (Clay) 6.5 m telescope. Twenty-one of our targets were observed twice, two or three days apart during 18--23 June, 2010 (UT100618--UT100623)\footnote{Throughout this paper we use a date format of UTYYMMDD.}, and then nearly a year later on 14--15 June, 2011 (UT110614--UT110615). Two targets were observed a third time on 9--10 May, 2012 (UT120509--UT120510), and another on 17 January, 2012 (UT120117). The complete log of our observations is recorded in Table~\ref{table_rv}. We used the red arm of the spectrograph and the $0.5\arcsec$ slit to obtain a spectrum from 4900 to 9150 \AA\ with a spectral resolution of $\sim$47,000. Total exposure times ranged from 2 to 30 minutes per target depending on stellar brightness to achieve a typical signal-to-noise ration (S/N) of $\approx$ 50 per resolution element at 8000 \AA. For the faintest stars, the S/N was roughly 10--20.

Data were reduced using the facility pipeline \citep{Kelson03}. Each stellar exposure was bias-subtracted and flat-fielded for pixel-to-pixel sensitivity variations. After optimal extraction, spectra were wavelength calibrated with a ThAr arc taken within an hour of the stellar exposure. To correct for instrumental drift, the telluric molecular oxygen A band (7619 -- 7657 \AA) was used to align the MIKE spectra to an average precision of 70 m~s$^{-1}$. We then corrected for the heliocentric velocity. \\

\section{Results}
\label{res}

\subsection{Radial Velocities}
\label{subsec:rvcalc}
Each target spectrum was cross-correlated against an RV standard star best matched to its SpT and observed on the same night. We used the IRAF\footnote{IRAF is distributed by the National Optical Astronomy Observatories, which are operated by the Association of Universities for Research in Astronomy, Inc., under a cooperative agreement with the National Science Foundation.} \textit{fxcor} routine \citep{Wyatt.85, Fitzpatrick.93} to obtain RVs and search for single- and double-lined spectroscopic binaries (SB1s and SB2s, respectively). The standards and their published RVs are listed in Table~\ref{table_stand}. We used orders 39,~40,~43,~44,~46,~48,~49,~51, and 53, which ranged from 6400 to 8900 \AA, for the cross-correlation to avoid telluric absorption lines and very low S/N regions. The RVs reported in Table~\ref{table_rv} are the averages of those measured for each order weighted by the average S/N in each order. Table~\ref{table_rv} also lists the standard star used, and \ion{Li}{1} and \ha EWs (see Appendix~\ref{age_accretion} for details of \ion{Li}{1} and \ha analysis). 

The uncertainties in our RV measurements include the standard error on the mean RV across the cross-correlated apertures (typically $\sim 400$ ms$^{-1}$), an uncertainty of 300 ms$^{-1}$ to account for RV distortions due to star spots on young stars \citep{Hatzes.02, Mohanty.02, Berdyugina.05, Desort.07},  and an additional 400 ms$^{-1}$ to account for the zero-point uncertainty in the RVs of our standard stars (see Table~\ref{table_stand}), all added in quadrature. In all cases, the systematic uncertainties from possible spot activity and the zero-point dominate the final uncertainties.  Note that $\Delta$(RV), the difference in RVs between epochs, need not include the uncertainty in the RV of the standards as the same standards were used.

The largest systematic uncertainties in these results will come from mismatched SpTs between the TD objects and the RV standard stars. While we endeavoured to match SpTs as closely as possible, the lack of G or K stars among our RV standards implies that for G and K-type TDs, systematics may remain unaccounted for. However, this should only affect SZ Cha (K0), SSTc2d  SSTc2d J162646.4-241160 (G5) and SR21A (G3--K?; a discussion of its SpT can be found in Appendix~\ref{sr21a}), since late K-type stars can be well matched by early M-type standards. Indeed, we see that the uncertainties on the RVs of SZ Cha are larger ($\sim 2\times$) than the average uncertainties.


The average RV of our Oph targets is $-6.6$ kms$^{-1}$ with a velocity dispersion of 1.3 kms$^{-1}$, which agrees well with other measurements for the RV of the cloud: $-5.64\pm$2.57 kms$^{-1}$ from \cite{Kurosawa.06} and $-6.27\pm$1.48 kms$^{-1}$ from \cite{Prato07}. There are no overlapping targets between our sample and those studies. 

Our average RV for Cha targets is 15$\pm$1 kms$^{-1}$, which is in good agreement with \cite{James.06}, who report an average of 12.8$\pm$3.6 kms$^{-1}$. However, our measurements of the average RV of CrA$+$Cor, -3$\pm$1 kms$^{-1}$,  differs by $\sim1.5\sigma$ from James et al.'s findings of $-$1.1$\pm$0.5 kms$^{-1}$. It is, however, in good agreement with \cite{Neuhauser.00}, who find an average RV of $-$2.6$\pm$1.4 kms$^{-1}$ for ROSAT-selected T Tauri stars in CrA.

The close agreement in mean RV values within our sample implies that any single RV measurement that deviates by more than 3$\sigma$ from the region average would be a highly probable RV-variable SB1 \citep{Kurosawa.06, Prato07}. We find no such targets in our sample.

\begin{deluxetable}{lcccccc}
\tabletypesize{\scriptsize}
\tablecaption{Observation log and measured properties of TD objects \label{table_rv}}
\tablewidth{0pt}
\tablehead{																					
\colhead{Target}        	&	              \colhead{Date observed}  	&	          \colhead{Standard}         	&	               \colhead{RV}    			&	       \colhead{Li I EW \tablenotemark{a}}                   			&	        \colhead{H$\alpha$ EW\tablenotemark{a}}                    			&	        \colhead{H$\alpha$ 10\%\tablenotemark{b}}                      			\\
\colhead{}              	&	              \colhead{dd/mm/yyyy}       	&	              \colhead{Star Used}       	&	              \colhead{kms$^{-1}$}                     			&	              \colhead{\AA}                                    			&	              \colhead{\AA}                                    			&	              \colhead{kms$^{-1}$}                                    	}		
\startdata                                                                                                                                                                           
SZ Cha\tablenotemark{c}	&	                     18/06/2010        	&	                     GJ 908    	&	12.1	$\pm$	0.5	&	0.36	$\pm$	0.04	&	-3.68	$\pm$	0.03	&	270	$\pm$	25	\\
        	&	                     21/06/2010        	&	                     GJ 908    	&	14.0	$\pm$	1.5	&	0.38	$\pm$	0.02	&	-26.6	$\pm$	0.5	&	479	$\pm$	47	\\
        	&	                     14/06/2011        	&	                     GJ 514    	&	17.0	$\pm$	1.3	&	0.42	$\pm$	0.05	&	-15.7	$\pm$	0.9	&	478	$\pm$	43	\\
        	&	                     mean      	&	               	&	14.0	$\pm$	2.0	&	0.39	$\pm$	0.07	&	-14	$\pm$	2	&	409	$\pm$	69	\\
T25\tablenotemark{c}               	&	                     18/06/2010        	&	                     GJ 908    	&	15.3	$\pm$	0.3	&	0.55	$\pm$	0.02	&	-10.7	$\pm$	0.8	&	327	$\pm$	29	\\
        	&	                     21/06/2010        	&	                     GJ 908    	&	15.6	$\pm$	0.3	&	0.58	$\pm$	0.02	&	-11	$\pm$	2	&	347	$\pm$	34	\\
        	&	                     14/06/2011        	&	                     GJ 514    	&	16.0	$\pm$	0.3	&	0.60	$\pm$	0.04	&	-21	$\pm$	1	&	444	$\pm$	49	\\
        	&	                     mean      	&	               	&	15.7	$\pm$	0.6	&	0.58	$\pm$	0.05	&	-15	$\pm$	1	&	373	$\pm$	66	\\
T35                     	&	                     21/06/2010        	&	                     GJ 908    	&	16.2	$\pm$	0.6	&	0.49	$\pm$	0.03	&	-118	$\pm$	15	&	482	$\pm$	45	\\
        	&	                     14/06/2011        	&	                     GJ 699    	&	16.8	$\pm$	0.6	&	0.41	$\pm$	0.08	&	-128	$\pm$	16	&	520	$\pm$	55	\\
        	&	                     mean      	&	               	&	16.5	$\pm$	0.8	&	0.45	$\pm$	0.09	&	-123	$\pm$	22	&	501	$\pm$	71	\\
RX J1852.3-3700\tablenotemark{d}                                	&	                     19/06/2010        	&	                     GJ 908    	&	-1.5	$\pm$	0.3	&	0.5	$\pm$	0.1	&	-45	$\pm$	6	&	320	$\pm$	30	\\
        	&	                     22/06/2010        	&	                     GJ 699    	&	-2.6	$\pm$	0.5	&	0.5	$\pm$	0.1	&	-33	$\pm$	2	&	325	$\pm$	33	\\
        	&	                     14/06/2011        	&	                     GJ 514    	&	-3.0	$\pm$	0.3	&	0.48	$\pm$	0.01	&	-27	$\pm$	3	&	326	$\pm$	35	\\
        	&	                     mean      	&	               	&	-2.4	$\pm$	0.6	&	0.5	$\pm$	0.2	&	-35	$\pm$	7	&	324	$\pm$	57	\\
CrA-4111                        	&	                     23/06/2010        	&	                     GJ 699    	&	-5.2	$\pm$	0.3	&	       	$<$	0.16      	&	-16	$\pm$	4	&	135	$\pm$	14	\\
G-49                    	&	                     15/06/2011        	&	                     GJ 699    	&	-2.7	$\pm$	0.3	&	0.51	$\pm$	0.04	&	-3	$\pm$	2	&	92	$\pm$	9	\\
        	&	                     09/05/2012        	&	                     GJ 699    	&	-2.6	$\pm$	0.3	&	0.62	$\pm$	0.01	&	-4.32	$\pm$	0.05	&	106	$\pm$	10	\\
        	&	                     10/05/2012        	&	                     GJ 699    	&	-2.8	$\pm$	0.3	&	0.62	$\pm$	0.01	&	-4.71	$\pm$	0.02	&	101	$\pm$	9	\\
        	&	                     mean      	&	               	&	-2.7	$\pm$	0.6	&	0.55	$\pm$	0.09	&	-4	$\pm$	2	&	100	$\pm$	16	\\
G-102           	&	                 15/06/2011    	&	                     GJ 699    	&	-2.8 $\pm$ 0.5		&	0.61	$\pm$	0.03	&	-14	$\pm$	4	&	189	$\pm$	18	\\
        	&	                     09/05/2012        	&	                     GJ 699    	&	-2.9	$\pm$	0.3	&	0.66	$\pm$	0.01	&	-20	$\pm$	2	&	153	$\pm$	14	\\
        	&	                     10/05/2012        	&	                     GJ 699    	&	-1.0	$\pm$	0.3	&	0.67	$\pm$	0.01	&	-17	$\pm$	1	&	152	$\pm$	7	\\
        	&	                     mean      	&	               	&	-2.2	$\pm$	0.7	&	0.65	$\pm$	0.03	&	-17	$\pm$	4	&	165	$\pm$	24	\\
\\                                                                                                                                                                                                                                                                                                                                                                                                                                                                                              																					
\hline                                                                                                                                                                                                                                                                                                                                                                                                                                                                                          																					
\\                                                                                                                                                                                                                                                                                                                                                                                                                                                                                                                                                                                              																					
SSTc2d J162118.5-225458\tablenotemark{c}             	&	                     21/06/2010        	&	                     GJ 908    	&	-6.3	$\pm$	0.4	&	0.48	$\pm$	0.01	&	-14.3	$\pm$	0.4	&	336	$\pm$	36	\\
        	&	                     22/06/2010        	&	                     GJ 908    	&	-6.6	$\pm$	0.4	&	0.5	$\pm$	0.01	&	-8.9	$\pm$	0.7	&	285	$\pm$	30	\\
        	&	                     mean      	&	               	&	-6.4	$\pm$	0.5	&	0.49	$\pm$	0.01	&	-11.6	$\pm$	0.8	&	310	$\pm$	47	\\
SSTc2d J162218.5-232148\tablenotemark{c}                 	&	                     19/06/2010        	&	                     GJ 908    	&	-7.9	$\pm$	0.4	&	0.4	$\pm$	0.2	&	-8.2	$\pm$	0.2	&	362	$\pm$	40	\\
        	&	                     21/06/2010        	&	                     GJ 908    	&	-9.7	$\pm$	0.4	&	0.4	$\pm$	0.1	&	-15	$\pm$	1	&	551	$\pm$	53	\\
        	&	                     14/06/2011        	&	                     GJ 514    	&	-7.5	$\pm$	0.4	&	0.42	$\pm$	0.03	&	-12.5	$\pm$	0.7	&	378	$\pm$	37	\\
        	&	                     mean      	&	               	&	-8.4	$\pm$	0.7	&	0.4	$\pm$	0.2	&	-12	$\pm$	2	&	430	$\pm$	76	\\
SSTc2d J162245.4-243124                 	&	                     20/06/2010        	&	                     GJ 699    	&	-5.3	$\pm$	0.3	&	0.5	$\pm$	0.4	&	-4.6	$\pm$	0.4	&	120	$\pm$	13	\\
        	&	                     22/06/2010        	&	                     GJ 699    	&	-5.4	$\pm$	0.3	&	0.51	$\pm$	0.01	&	-5.1	$\pm$	0.8	&	124	$\pm$	12	\\
        	&	                     14/06/2011        	&	                     GJ 699    	&	-4.0	$\pm$	0.3	&	0.55	$\pm$	0.05	&	-4.2	$\pm$	0.8	&	115	$\pm$	4	\\
        	&	                     mean      	&	               	&	-4.8	$\pm$	0.6	&	0.5	$\pm$	0.4	&	-5	$\pm$	1	&	120	$\pm$	18	\\
SSTc2d J162309.2-241705                	&	                     20/06/2010        	&	                     GJ 908    	&	-3.8	$\pm$	0.4	&	0.28	$\pm$	0.06	&	-17.3	$\pm$	0.5	&	419	$\pm$	46	\\
        	&	                     22/06/2010        	&	                     GJ 908    	&	-2.9	$\pm$	0.4	&	0.29	$\pm$	0.01	&	-16.3	$\pm$	0.8	&	365	$\pm$	30	\\
        	&	                     14/06/2011        	&	                     GJ 699    	&	-3.6	$\pm$	0.4	&	0.29	$\pm$	0.01	&	-14	$\pm$	1	&	363	$\pm$	35	\\
        	&	                     mean      	&	               	&	-3.4	$\pm$	0.7	&	0.29	$\pm$	0.06	&	-16	$\pm$	2	&	382	$\pm$	65	\\
SSTc2d J162332.8-225847                 	&	                     20/06/2010        	&	                     GJ 699    	&	-6.9	$\pm$	0.3	&	0.54	$\pm$	0.01	&	-13	$\pm$	3	&	153	$\pm$	15	\\
        	&	                     22/06/2010        	&	                     GJ 699    	&	-7.0	$\pm$	0.3	&	0.44	$\pm$	0.03	&	-11	$\pm$	3	&	122	$\pm$	13	\\
        	&	                     mean      	&	               	&	-6.9	$\pm$	0.4	&	0.49	$\pm$	0.04	&	-12	$\pm$	4	&	137	$\pm$	20	\\
SSTc2d J162336.1-240221                 	&	                     20/06/2010        	&	                     GJ 699    	&	-6.7	$\pm$	0.3	&	              	$<$	0.28	&	-12	$\pm$	3	&	93	$\pm$	8	\\
        	&	                     22/06/2010        	&	                     GJ 699    	&	-7.3	$\pm$	0.3	&	0.36	$\pm$	0.01	&	-8	$\pm$	1	&	113	$\pm$	11	\\
        	&	                     mean      	&	               	&	-7.0	$\pm$	0.4	&	0.4	$\pm$	0.3	&	-10	$\pm$	3	&	103	$\pm$	14	\\
SSTc2d J162506.9-235050\tablenotemark{c}                 	&	                     20/06/2010        	&	                     GJ 699    	&	-6.5	$\pm$	0.3	&	0.51	$\pm$	0.01	&	-9	$\pm$	1	&	259	$\pm$	26	\\
        	&	                     22/06/2010        	&	                     GJ 699    	&	-6.9	$\pm$	0.3	&	0.53	$\pm$	0.01	&	-12	$\pm$	3	&	341	$\pm$	34	\\
        	&	                     14/06/2011        	&	                     GJ 699    	&	-7.0	$\pm$	0.3	&	0.57	$\pm$	0.01	&	-14	$\pm$	4	&	383	$\pm$	48	\\
        	&	                     mean      	&	               	&	-6.8	$\pm$	0.6	&	0.53	$\pm$	0.01	&	-12	$\pm$	5	&	328	$\pm$	64	\\
SSTc2d J162623.7-244314                 	&	                     21/06/2012        	&	                     GJ 908    	&	-7.2	$\pm$	0.4	&	0.5	$\pm$	0.2	&	-4.5	$\pm$	0.4	&	239	$\pm$	25	\\
        	&	                     14/06/2011        	&	                     GJ 908    	&	-6.7	$\pm$	0.4	&	0.56	$\pm$	0.02	&	-11	$\pm$	1	&	467	$\pm$	47	\\
        	&	                     mean      	&	               	&	-6.9	$\pm$	0.6	&	0.5	$\pm$	0.2	&	-8	$\pm$	1	&	353	$\pm$	53	\\
SSTc2d J162646.4-241160                 	&	                     20/06/2010        	&	                     GJ 908    	&	-6.3	$\pm$	0.4	&	0.5	$\pm$	0.1	&	-11.7	$\pm$	0.4	&	346	$\pm$	31	\\
        	&	                     22/06/2010        	&	                     GJ 908    	&	-6.6	$\pm$	0.3	&	0.4	$\pm$	0.2	&	-14.7	$\pm$	0.8	&	368	$\pm$	36	\\
        	&	                     14/06/2011        	&	                     GJ 514    	&	-6.2	$\pm$	0.4	&	0.4	$\pm$	0.1	&	-17	$\pm$	2	&	356	$\pm$	34	\\
        	&	                     mean      	&	               	&	-6.4	$\pm$	0.7	&	0.4	$\pm$	0.3	&	-14.38	$\pm$	2.2	&	357	$\pm$	58	\\
DoAr28                  	&	                     20/06/2010        	&	                     GJ 908    	&	-8.6	$\pm$	0.3	&	0.45	$\pm$	0.01	&	-25	$\pm$	2	&	451	$\pm$	40	\\
        	&	                     22/06/2010        	&	                     GJ 908    	&	-8.0	$\pm$	0.4	&	0.4	$\pm$	0.2	&	-29	$\pm$	2	&	441	$\pm$	47	\\
        	&	                     14/06/2011        	&	                     GJ 514    	&	-7.9	$\pm$	0.3	&	0.4	$\pm$	0.2	&	-28	$\pm$	1	&	436	$\pm$	40	\\
        	&	                     mean      	&	               	&	-8.2	$\pm$	0.6	&	0.4	$\pm$	0.3	&	-27	$\pm$	3	&	443	$\pm$	74	\\
SR21A                   	&	                     14/06/2011        	&	                     GJ 908    	&	-3.6	$\pm$	0.5	&	0.13	$\pm$	0.02	&	1.1	$\pm$	   0.15\tablenotemark{f}                                   	&	                                       	\nodata		\\
SSTc2d J162738.3-235732                 	&	                     19/06/2010        	&	                     GJ 908    	&	-7.4	$\pm$	0.3	&	0.47	$\pm$	0.01	&	-16	$\pm$	2	&	291	$\pm$	33	\\
        	&	                     22/06/2010        	&	                     GJ 908    	&	-6.5	$\pm$	0.3	&	0.49	$\pm$	0.02	&	-14	$\pm$	3	&	275	$\pm$	29	\\
        	&	                     14/06/2011        	&	                     GJ 514    	&	-6.4	$\pm$	0.3	&	0.5	$\pm$	0.05	&	-13.4	$\pm$	0.9	&	292	$\pm$	29	\\
        	&	                     mean      	&	               	&	-6.8	$\pm$	0.6	&	0.49	$\pm$	0.05	&	-15	$\pm$	4	&	286	$\pm$	53	\\
SSTc2d J162739.0-235818\tablenotemark{c}                 	&	                     19/06/2010        	&	                     GJ 699    	&	-7.7	$\pm$	0.5	&	0.49	$\pm$	0.01	&	-22	$\pm$	1	&	373	$\pm$	37	\\
        	&	                     21/06/2010        	&	                     GJ 699    	&	-7.3	$\pm$	0.7	&	0.52	$\pm$	0.01	&	-13	$\pm$	4	&	282	$\pm$	28	\\
        	&	                     14/06/2011        	&	                     GJ 514    	&	-7.0	$\pm$	0.4	&	0.54	$\pm$	0.01	&	-9.9	$\pm$	0.2	&	261	$\pm$	27	\\
        	&	                     mean      	&	               	&	-7.3	$\pm$	0.9	&	0.51	$\pm$	0.02	&	-15	$\pm$	4	&	306	$\pm$	54	\\
SSTc2d J162740.3-242204                 	&	                     22/06/2010        	&	                     GJ 908    	&	-6.0	$\pm$	0.4	&	0.5	$\pm$	0.1	&	-10.3	$\pm$	0.2	&	289	$\pm$	29	\\
        	&	                     14/06/2011        	&	                     GJ 514    	&	-7.2	$\pm$	0.3	&	0.51	$\pm$	0.01	&	-10.77	$\pm$	0.09	&	362	$\pm$	32	\\
        	&	                     mean      	&	               	&	-6.6	$\pm$	0.5	&	0.5	$\pm$	0.1	&	-10.5	$\pm$	0.2	&	325	$\pm$	43	\\
SSTc2d J162802.6-235504                 	&	                     20/06/2010        	&	                     GJ 908    	&	-7.3	$\pm$	0.4	&	0.51	$\pm$	0.04	&	              	$<$	1	&		\nodata		\\
        	&	                     22/06/2010        	&	                     GJ 908    	&	-6.6	$\pm$	0.3	&	0.57	$\pm$	0.08	&	-3.7	$\pm$	0.2	&	77	$\pm$	6	\\
        	&	                     mean      	&	               	&	-6.9	$\pm$	0.5	&	0.54	$\pm$	0.09	&	-3.5	$\pm$	0.9	&	77	$\pm$	6	\\
SSTc2d J162821.5-242155                 	&	                     21/06/2010        	&	                     GJ 699    	&	-7.1	$\pm$	0.3	&	0.4	$\pm$	0.2	&	-2.9	$\pm$	0.6	&	82	$\pm$	7	\\
        	&	                     22/06/2010        	&	                     GJ 699    	&	-7.5	$\pm$	0.3	&	0.4	$\pm$	0.2	&	-3	$\pm$	1	&	95	$\pm$	5	\\
        	&	                     mean      	&	               	&	-7.3	$\pm$	0.5	&	0.4	$\pm$	0.3	&	-3	$\pm$	1	&	88	$\pm$	9	\\
SSTc2d J162854.1-244744                 	&	                     20/06/2010        	&	                     GJ 908    	&	-5.0	$\pm$	0.3	&	0.58	$\pm$	0.02	&	-21.7	$\pm$	0.4	&	401	$\pm$	37	\\
        	&	                     22/06/2010        	&	                     GJ 908    	&	-5.1	$\pm$	0.3	&	0.59	$\pm$	0.02	&	-24	$\pm$	6	&	314	$\pm$	29	\\
        	&	                     14/06/2011        	&	                     GJ 514    	&	-4.4	$\pm$	0.3	&	0.6	$\pm$	0.3	&	-65	$\pm$	13	&	423	$\pm$	52	\\
        	&	                     mean      	&	               	&	-4.8	$\pm$	0.6	&	0.6	$\pm$	0.3	&	-37	$\pm$	14	&	379	$\pm$	70	\\
SSTc2d J163020.0-233108                 	&	                     21/06/2010        	&	                     GJ 908    	&	-7.2	$\pm$	0.4	&	0.57	$\pm$	0.75	&	-2.1	$\pm$	0.4	&	87	$\pm$	7	\\
        	&	                     22/06/2010        	&	                     GJ 699    	&	-6.7	$\pm$	0.3	&	0.56	$\pm$	0.57	&	-1.6	$\pm$	0.4	&	83	$\pm$	7	\\
        	&	                     mean      	&	               	&	-7.0	$\pm$	0.5	&	0.57	$\pm$	0.94	&	-1.8	$\pm$	0.6	&	85	$\pm$	10	\\
SSTc2d J163033.9-242806                 	&	                     20/06/2010        	&	                     GJ 699    	&	-6.5	$\pm$	0.5	&	1.3	$\pm$	0.2	&	-19	$\pm$	3	&	332	$\pm$	33	\\
        	&	                     22/06/2010        	&	                     GJ 699    	&	-6.1	$\pm$	0.4	&	0.51	$\pm$	0.08	&	-17	$\pm$	1	&	247	$\pm$	24	\\
        	&	                     mean      	&	               	&	-6.3	$\pm$	0.6	&	0.9	$\pm$	0.2	&	-18	$\pm$	4	&	290	$\pm$	41	\\
SSTc2d J163145.4-244307            	&	                     21/06/2010        	&	                     GJ 699    	&	-5.0	$\pm$	0.4	&	0.5	$\pm$	0.4	&	-12.2	$\pm$	0.2	&	455	$\pm$	49	\\
        	&	                     22/06/2010        	&	                     GJ 699    	&	-2.4	$\pm$	0.4	&	0.3	$\pm$	0.2	&	-37	$\pm$	2	&	446	$\pm$	39	\\
        	&	                     14/06/2011        	&	                     GJ 699    	&	-4.8	$\pm$	0.3	&	0.5	$\pm$	0.05	&	-60	$\pm$	9	&	364	$\pm$	37	\\
        	&	                     mean      	&	               	&	-4.1	$\pm$	0.7	&	0.4	$\pm$	0.4	&	-36	$\pm$	9	&	422	$\pm$	73	\\
SSTc2d J163154.4-250349         	&	                     21/06/2010        	&	                     GJ 699    	&	-4.6	$\pm$	0.4	&	0.3	$\pm$	0.1	&	-106	$\pm$	29	&	374	$\pm$	43	\\
        	&	                     14/06/2011        	&	                     GJ 514    	&	-3.7	$\pm$	0.3	&	0.34	$\pm$	0.07	&	-139	$\pm$	33	&	402	$\pm$	35	\\
        	&	                     mean      	&	               	&	-4.1	$\pm$	0.5	&	0.3	$\pm$	0.1	&	-123	$\pm$	44	&	388	$\pm$	56	\\
SSTc2d J163154.7-250324 (SB2)\tablenotemark{e}                          	&	                     20/06/2010        	&	                     GJ 699    	&	-12.1	$\pm$	1.0	&	0.56	$\pm$	0.01	&	-8.4	$\pm$	0.2	&	394	$\pm$	38	\\
        	&	                     22/06/2010        	&	                     GJ 699    	&	-16.4	$\pm$	1.2	&	0.5	$\pm$	0.2	&	-6	$\pm$	1	&	301	$\pm$	27	\\
        	&	                     14/06/2011        	&	                     GJ 514        	&	-5.4	$\pm$	0.5	&	0.5	$\pm$	0.2	&	-7	$\pm$	2	&	397	$\pm$	39	\\
        	&	                     mean      	&	               	&	-11	$\pm$	2	&	0.5	$\pm$	0.2	&	-7	$\pm$	2	&	364	$\pm$	61	\\
SSTc2d J163205.5-250236                 	&	                     21/06/2010        	&	                     GJ 908    	&	-6.7	$\pm$	0.3	&	0.50	$\pm$	0.09	&	-34	$\pm$	2	&	461	$\pm$	43	\\
        	&	                     17/01/2012        	&	                     GJ 908    	&	-6.0	$\pm$	0.4	&	0.51	$\pm$	0.01	&	-25	$\pm$	2	&	427	$\pm$	39	\\
        	&	                     mean      	&	               	&	-6.4	$\pm$	0.5	&	0.51	$\pm$	0.09	&	-29	$\pm$	3	&	444	$\pm$	58	\\
SSTc2d J163355.6-244205                 	&	                     22/06/2010        	&	                     GJ 699    	&	-6.1	$\pm$	0.4	&	0.52	$\pm$	0.02	&	-2.7	$\pm$	0.6	&	255	$\pm$	23	\\
        	&	                     14/06/2011        	&	                     GJ 514    	&	-5.8	$\pm$	0.3	&	0.53	$\pm$	0.02	&	-7	$\pm$	3	&	283	$\pm$	31	\\
        	&	                     mean      	&	               	&	-6.0	$\pm$	0.5	&	0.53	$\pm$	0.03	&	-5	$\pm$	3	&	269	$\pm$	39	\\
\enddata        																						

\tablenotetext{a}{If the spectral line could not be distinguished from the continuum, a 2$\sigma$ limit is shown.}
								
\tablenotetext{b}{\ha 10\% velocity width \citep{WhiteBasri.03} variability is due to stellar activity or variable accretion \citep[e.g.][]{Jaywardhana.03, Natta.04}. }		

\tablenotetext{c}{These targets were found to be active stars or variable accretors and are discussed in greater detail in Appendix~\ref{age_accretion}.}

\tablenotetext{d}{\cite{White07} measured an RV of $-$1.46$\pm$2.21 kms$^{-1}$, consistent with our measurements.}
																							
\tablenotetext{e}{The RV listed here is the systemic velocity of the binary system, assuming mass ratio $q=0.95$ (See Section~\ref{results}). The EWs listed are the blended values at each epoch.}																						
\tablenotetext{f}{Note that SR21A shows \ha\ in absorption rather than in emission. See Appendix~\ref{sr21a} for further discussion of this object.}																								
\end{deluxetable}																						
\clearpage

\begin{deluxetable}{lcc}
\tabletypesize{\scriptsize}
\tablecaption{Radial Velocity Standards\label{table_stand}}
\tablewidth{0pt}
\tablehead{
\colhead{Name}       &       \colhead{Spectral Type}    &       \colhead{RV\tablenotemark{a}} \\
\colhead{}	&\colhead{}	&\colhead{kms$^{-1}$}
}
\startdata
GJ 514 & M1 & 14.56$\pm$0.40\\
GJ 699 & M4 & -110.51$\pm$0.40\\
GJ 908 & M2 & -71.15$\pm$0.40\\
\enddata
\tablenotetext{a}{RVs are taken from \cite{Nidever.02}, who quote an RV stability of 0.05 kms$^{-1}$. The zero point of the absolute RVs in this study is uncertain at the 0.4 kms$^{-1}$ level \citep{Marcy.89}.}
\end{deluxetable}
\clearpage

\subsection{Rotational Velocities}
\label{subsec:vsinicalc}
We determined rotational velocities ($v\sin i$) following the `Fourier Method' \citep[e.g.][]{Carroll.33, Gray.76, Reiners.01, Gray.05, Simon-Diaz.07}. 
Using telluric lines, we determined a small amount of instrumental broadening ($\sim0.2$\AA), which we absorbed into model lines of zero rotation.
We extracted the excess broadening due to rotation using the relatively high S/N \ion{Li}{1} doublet at 6708 \AA\,  and \ion{Ca}{1} at 6718 \AA. While forbidden atomic transitions such as [\ion{Fe}{3}] do not run the risk of confusing rotational broadening for thermal broadening, their S/N were consistently too low for use.

Our results were sensitive to the assumed limb-darkening coefficient, $\epsilon$, with an $\sim15$\, kms$^{-1}$ deviation in $v\sin i$ for a change $\Delta\epsilon\sim 0.1$. Due to this sensitivity, we used the \cite{Claret.00} linear limb-darkening coefficient catalog for each source, using characteristic values of $\log g$ and effective temperatures \citep{Gizis.97, Casagrande.08, Raj.13} for their literature SpTs (for G and K stars) and those determined by their TiO-7140 indices (for M stars; see Section~\ref{subsec:calcSpTs}). 

For each observation, we took a S/N-weighted average of the $v\sin i$ as measured from the \ion{Li}{1} and \ion{Ca}{1} lines. For each target, we took the average of the per-observation measurements. Uncertainties were added in quadrature. Two targets in our sample have $v\sin i$ measurements reported in the literature: RX J1852.3-3700 \citep[23.05$\pm$3.59\,kms$^{-1}$;][]{White.07} and SSTc2d J162740.3-242204 \citep[14.7$\pm$0.9\,kms$^{-1}$;][]{Torres.06}. Both values are within the uncertainties of our measured values. We present our measurements in Table~\ref{table_extra}.

\subsection{SpT measurements}
\label{subsec:calcSpTs}

\begin{figure}[h]
\centering
\includegraphics[scale=0.5]{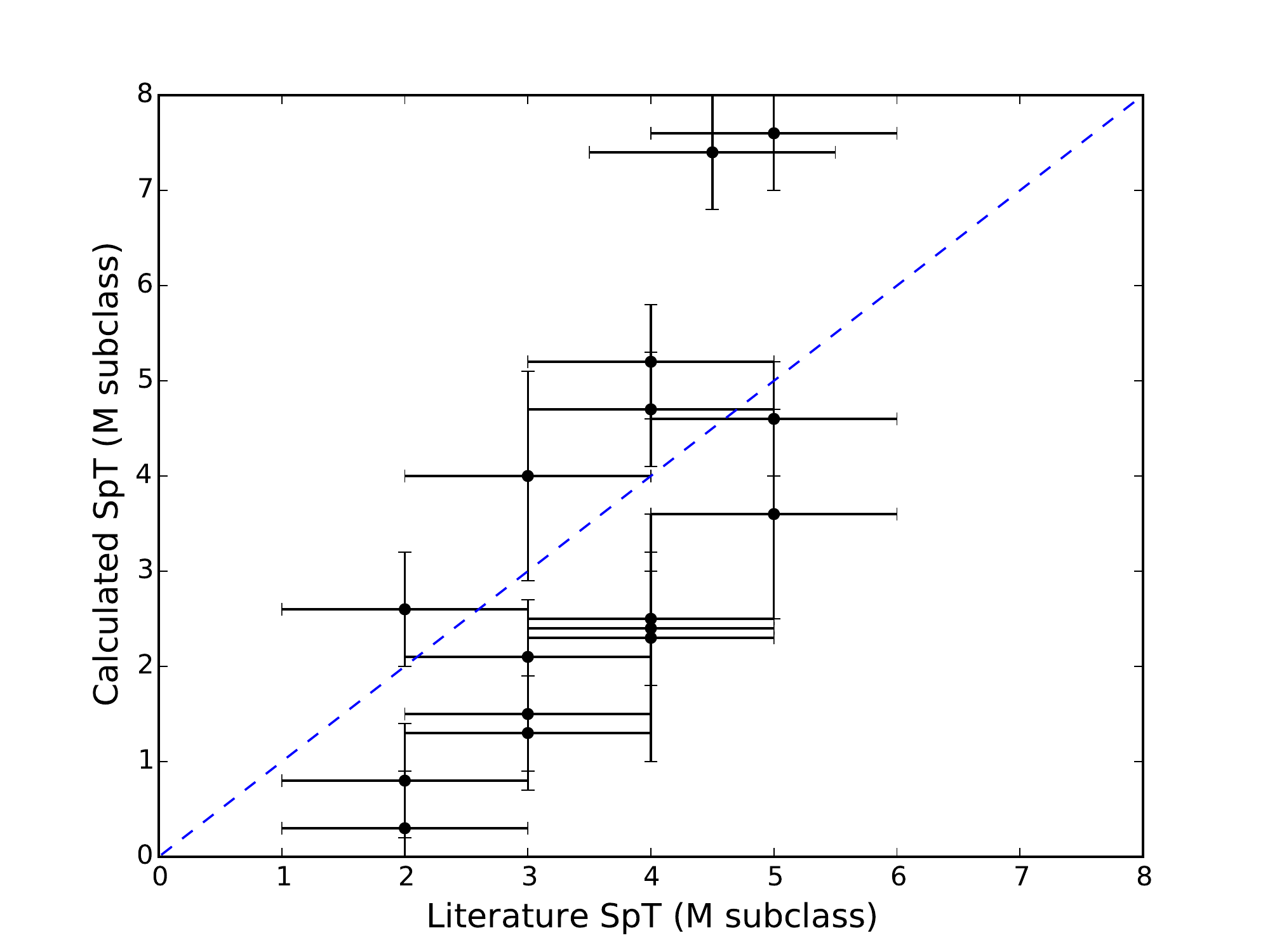}
\caption{The M subclasses of targets reported as M dwarfs in the literature (i.e. 0$\rightarrow$M0, 1$\rightarrow$M1, etc.), with their literature subclasses compared with our own measurements based on their TiO-7140 indicies. A 1:1 relation is overlaid in blue. Our measurements are broadly consistent with the literature values. Larger deviations from the literature values at subclasses $\geq5$ occur because the TiO-7140 index is less well-constrained for late M SpTs.}
\label{fig:SpTcalc}
\end{figure}

M dwarf spectra are dominated by TiO molecular bands. \cite{Wilking.05} defined the TiO-7140 index as a way of determining the SpTs of M dwarfs: a ratio of the mean flux of two 50\AA\,bands centered on 7035\AA\,(the continuum band) and 7140\AA\,(the TiO band). We used the calibration given in \cite{Shkolnik.09}

\begin{equation}
\rm{SpT} = (\TiOindex - 1.0911)/0.1755
\end{equation}

\noindent to measure the SpTs of all of our targets. Strictly, this relation holds for SpTs M0--M5, which covers the range of SpT literature values of M dwarfs in our sample\footnote{An interesting case is that of target SSTc2d J162309.2-241705, for which C10 report an uncertain spectral type of ``M?". We measure a negative M subclass for this target, suggesting that it may be a late K.}.
Our results are shown in Figure~\ref{fig:SpTcalc}. The error bars for literature values are $\pm1$ subclass, largely from C10. The error bars for our own measurements are the root mean square=0.6 scatter of the \cite{Shkolnik.09} calibration, added in quadrature with the standard deviation of the measured SpTs across the multiple epochs each target was observed. They favor a linear fit of

\begin{equation}
\rm{SpT}_{\rm{This\,work}} = (1.5\pm0.4)\rm{SpT}_{\rm{Lit.}} + (-2 \pm 1)
\end{equation}

\noindent i.e. we measure SpTs consistent with the literature values to $\sim\pm1$ SpT subclass. Our measured SpT per target is shown in Table~\ref{table_extra}, along with the literature values (see references in Table~\ref{table_td}) and $v\sin i$ (see above). We do not revise K-type SpTs with the exception of SSTc2d  J163154.7-250324, which we find to be an SB2 (see Section~\ref{results}). For this system we report the derived SpTs and $v\sin i$ for each component. The lack of TiO in the spectra of the two G stars in our sample (SSTc2d  J162646.4-241160 and SR21A) indicates that they are hotter than M stars, but the difference between very young G and K stars not well-constrained \citep[e.g.][]{Beuther.14}, and therefore we do not revise their SpTs.


\begin{deluxetable}{llll}
\tabletypesize{\scriptsize}
\tablecaption{Spectral types and rotational velocities\label{table_extra}}
\tablewidth{0pt}
\tablehead{											
\colhead{Name}	&	\colhead{SpT}	&	\colhead{SpT}			&	\colhead{$v\sin i$}			\\
	&	\colhead{(Lit.)}	&	\colhead{(This work)}			&	\colhead{kms$^{-1}$}			
}											
\startdata											
SZ Cha                  	&	              K0       	&				&	34.0	$\pm$	0.4	\\
T25                     	&	              M2.5       	&	M1.4	$\pm$	0.6	&	24.3	$\pm$	0.4	\\
T35                     	&	              K8       	&				&	29	$\pm$	3	\\
RX J1852.3-3700                 	&	              K7       	&				&	19.5	$\pm$	0.5\tablenotemark{a}	\\
CrA-4111                        	&	              M4.5     	&	M7.4	$\pm$	0.6	&	65	$\pm$	6	\\
G-49                    	&	              M4       	&	M5.2	$\pm$	0.6	&	43	$\pm$	4	\\
G-102                   	&	              M5       	&	M7.6	$\pm$	0.6	&	41	$\pm$	4	\\
\\\hline\\											
SSTc2d  J162118.5-225458                	&	              M2       	&	M2.6	$\pm$	0.6	&	40	$\pm$	3	\\
SSTc2d  J162218.5-232148                	&	              K5       	&				&	35	$\pm$	16	\\
SSTc2d  J162245.4-243124                	&	              M3       	&	M2.1	$\pm$	0.6	&	42.3	$\pm$	0.2	\\
SSTc2d J162309.2-241705         	&	              M?       	&		K?		&	45.0	$\pm$	0.9	\\
SSTc2d  J162332.8-225847                	&	              M5       	&	M4.6	$\pm$	0.6	&	49	$\pm$	4	\\
SSTc2d  J162336.1-240221                	&	              M5       	&	M3.6	$\pm$	1.1	&	64	$\pm$	5	\\
SSTc2d  J162506.9-235050                	&	              M3       	&	M1.3	$\pm$	0.6	&	46	$\pm$	6	\\
SSTc2d  J162623.7-244314                	&	              K5       	&				&	10	$\pm$	1	\\
SSTc2d  J162646.4-241160                	&	              G5       	&				&	49	$\pm$	5	\\
DoAr28                  	&	              K?       	&				&	27	$\pm$	19	\\
SR21A                   	&	              G3       	&				&	65	$\pm$	3	\\
SSTc2d  J162738.3-235732                	&	              K5       	&				&	42	$\pm$	2	\\
SSTc2d  J162739.0-235818                	&	              K6       	&				&	43	$\pm$	23	\\
SSTc2d  J162740.3-242204                	&	              K5       	&				&	18	$\pm$	2\tablenotemark{b}	\\
SSTc2d  J162802.6-235504                	&	              M3       	&	M4.0	$\pm$	1.1	&	60	$\pm$	12	\\
SSTc2d  J162821.5-242155                	&	              M3       	&	M1.5	$\pm$	0.6	&	63	$\pm$	6	\\
SSTc2d  J162854.1-244744                	&	              M2       	&	M0.3	$\pm$	0.6	&	49	$\pm$	7	\\
SSTc2d  J163020.0-233108                	&	              M4       	&	M2.4	$\pm$	0.6	&	50	$\pm$	6	\\
SSTc2d  J163033.9-242806                	&	              M4       	&	M4.7	$\pm$	0.6	&	60	$\pm$	8	\\
SSTc2d  J163145.4-244307                	&	              M4       	&	M2.3	$\pm$	1.3	&	57	$\pm$	12	\\
SSTc2d  J163154.4-250349                	&	              M4       	&	M2.5	$\pm$	0.7	&	57	$\pm$	14	\\
SSTc2d  J163154.7-250324A\tablenotemark{c}	               &	              K7       	&		K7.0$\pm$0.5		&	18.2	$\pm$	0.9\\
SSTc2d  J163154.7-250324B\tablenotemark{c}               &	                     	    &	 K9.0$\pm$0.5			&	23	$\pm$	5	\\
SSTc2d  J163205.5-250236                	&	              M2       	&	M0.8	$\pm$	0.6	&	48	$\pm$	5	\\
SSTc2d  J163355.6-244205                	&	              K7       	&				&	43	$\pm$	4	\\
\enddata																					
\tablenotetext{a}{\protect\cite{White.07} report a $v\sin i = 23.05\pm3.59$ kms$^{-1}$ for this target, within 1$\sigma$ of our measured value.}	
\tablenotetext{b}{\protect\cite{Torres.06} report a $v\sin i = 14.7\pm0.9$ kms$^{-1}$ for this target, within 2$\sigma$ of our measured value.}		
\tablenotetext{c}{This target is an SB2. In this Table we report the properties of each component. See Section~\ref{results}.}							
\end{deluxetable}

\section{Spectroscopic Binary Detection}
\label{results}

Single-lined SBs, SB1s are revealed by significant RV variability between observations. We performed $\chi^{2}$ tests on the RVs of targets with single-peaked cross-correlation functions (CCFs) using the target's average RV as a flat prior and found no SB1s in our sample of 31 TD objects. As mentioned above, the fact that all of the RVs agree with both the sample average for their SFR and the association average from the literature argues against any of our targets being long-period SB1s, or $P\sim1$\,yr systems whose RV was serendipitously measured by us at the same orbital phase each year.

Cross-correlation of the target spectra with an RV standard spectrum reveals whether or not a star is an SB2 if the orbital phase at the time of observation allows resolvable RV motion such that CCF is double-peaked.  We found one SB2 in the sample (SSTc2d J163154.7-250324), which we discuss in the subsections below.

\subsection{Measured Properties of SSTc2d J163154.7-250324}

\begin{figure}
\centering
\includegraphics[scale=0.5]{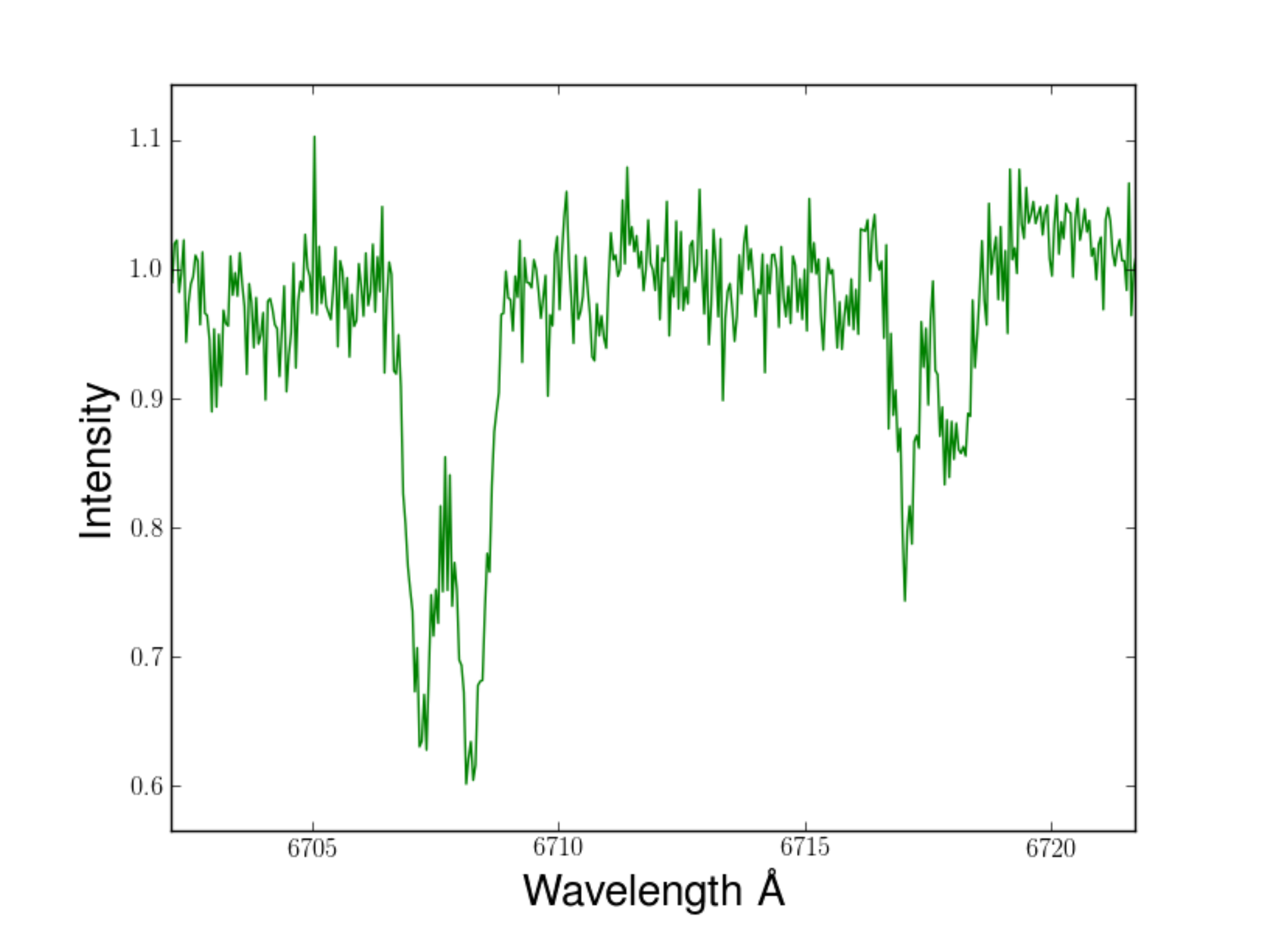}
\caption{The spectrum of SSTc2d J163154.7-250324 on UT110614. The SB2 nature of the system is evident in all absorption features, including the Li lines at 6708 \AA.}
\label{fig_SB2_LiI}
\end{figure}

\begin{figure}
\centering
\includegraphics[scale=0.5]{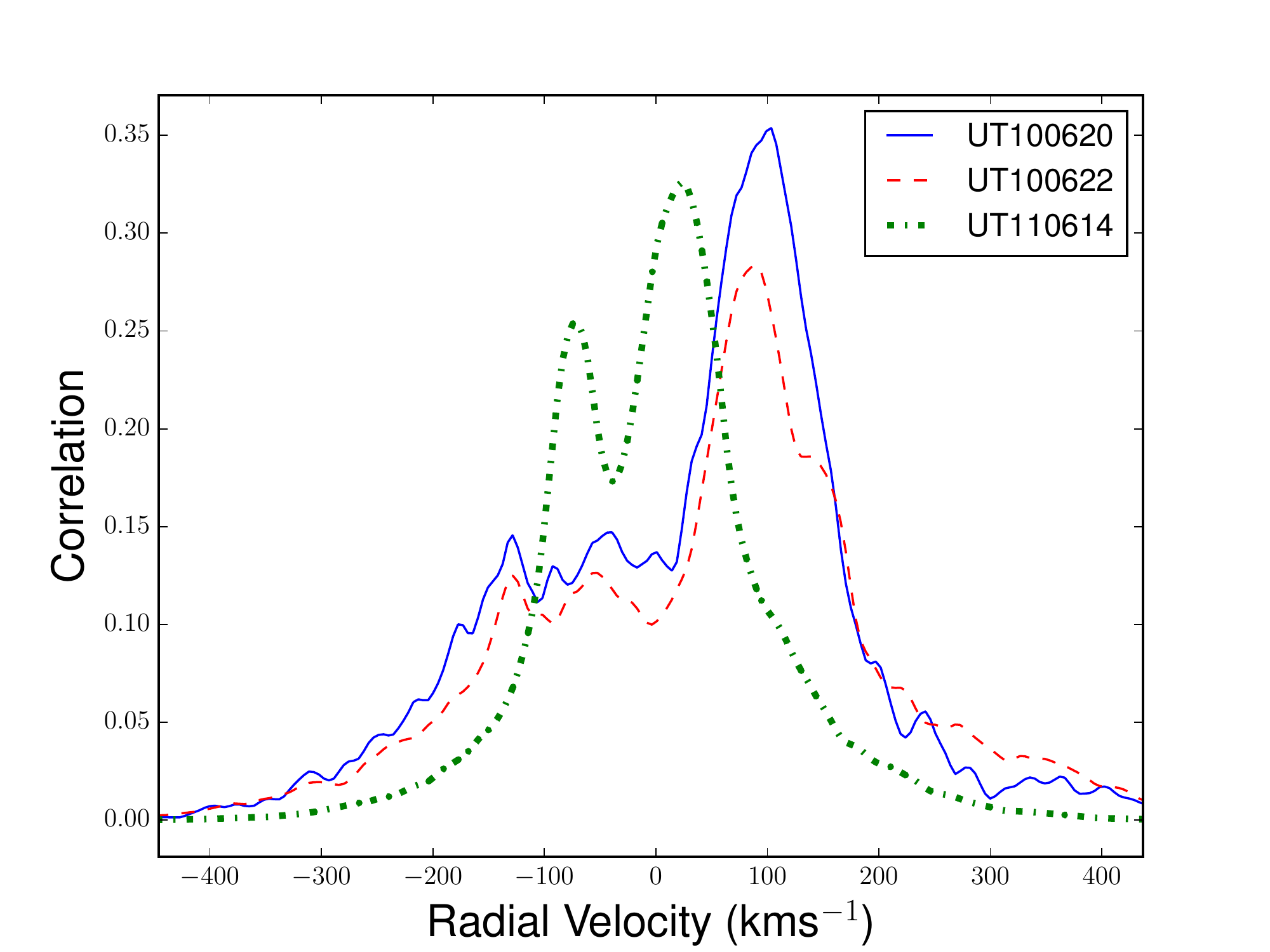}
\caption{The CCFs of SSTc2d J163154.7-250324 averaged across apertures used. Spectra taken in 2010 were cross-correlated against GJ 699 and those in 2011 against GJ 514.}
\label{fig_SB2_CCF}
\end{figure}

The stellar features were blended enough on UT100620 and UT100622 that we could not confirm it as an SB2 using those observations alone. SSTc2d J163154.7-250324 clearly exhibited a double-peaked CCF one year later on UT110614. We measured the RVs of the primary and secondary components to be 16$\pm$2 kms$^{-1}$ and -27$\pm$2 kms$^{-1}$, respectively. This gives a systemic RV of $\gamma$=-5.4$\pm$0.5 kms$^{-1}$ based on fluxes in the two CCF peaks (flux ratio of 0.68$\pm$0.07; see below). We measured values of $v\sin i$ for the primary and secondary as 18.2$\pm$0.9 kms$^{-1}$ and 23$\pm$5 kms$^{-1}$, respectively, following the method described in Section~\ref{subsec:vsinicalc} on the unblended absorption lines seen on UT110614.
A section of the stellar spectrum and the CCFs of this SB2 are shown in Figures~\ref{fig_SB2_LiI} and~\ref{fig_SB2_CCF} respectively.  

Since we resolved the two CCF peaks, we were able to estimate spectral types and component masses of the individual stars. Assuming a flux-weighted relation (where $f_i$ is the integrated flux of Gaussian fits to the cross-correlation peaks at R band wavelengths; Figure~\ref{fig_SB2_CCF}) between component and integrated SpTs \citep{Cruz.02, Reid.02, Daemgen.07} for primary star $A$ and secondary $B$:

\begin{equation}\label{eq:SpT_flux}
\mbox{SpT}_{\rm int} = (f_A~\mbox{SpT}_A + f_B~\mbox{SpT}_B)/(f_A + f_B)
\end{equation}

Using the Kepler's Third law we related component masses $M_n$ to velocity amplitudes $K_n$:
\begin{equation}\label{eq:Mass_Vel}
M_A/M_B = K_A/K_B
\end{equation}

We imposed a limit on SpT and magnitude \citep{Daemgen.07, Shkolnik.10}:
\begin{equation}\label{eq:SpT_mag}
\Delta{\rm R} = M_R(\mbox{SpT}_B)-M_R(\mbox{SpT}_A)
\end{equation}

We measure a flux ratio of 0.68$\pm$0.07 for the CCF peaks. Thus, we determine that SSTc2d J163154.7-250324 is composed of a K7($\pm$0.5) and a K9($\pm$0.5) star, with mass ratio $q\simeq$0.95. Using Kepler's Third Law, we calculated the orbital limits to be $a<$0.6 AU and period $P<$150 days. Our measurements of this system are summarized in Table~\ref{table_SB2props}.

\begin{deluxetable}{lccc}
\tabletypesize{\scriptsize}
\tablecaption{Properties of the SSTc2d J163154.7-250324 system\label{table_SB2props}}
\tablewidth{0pt}
\tablehead{
\colhead{Property}       &       \colhead{System}    &       \colhead{A} & \colhead{B} \\
}
\startdata
SpT & K7\tablenotemark{a} & K7.0$\pm$0.5 & K9.0$\pm$0.5 \\
RV (kms$^{-1}$) & -5.4$\pm$0.5 & 16$\pm$2 & -27$\pm$2 \\
$v\sin i$ (kms$^{-1}$) &  & 18.2$\pm$0.9 & 23$\pm$5\\
Separation (AU) & $<$0.6 & & \\
Period (days) & $<$150 & & \\
Mass ratio & 0.95 & & \\
\enddata
\tablenotetext{a}{\protect\cite{Cieza.10}}
\end{deluxetable}

\subsection{Comparison to the literature}

Before the SB2 SSTc2d J163154.7-250324 was identified as a TD, it had already been measured to have an infrared excess in its SED by 2MASS and exhibit the visible spectral properties of a YSO \citep[][who refer to the system as H$\alpha$74 and/or ISO-Oph 207]{Ratzka.05}. The disk is unresolved in the CHARM2 catalog \citep{Richichi.05}. \cite{Padgett.08} reported the YSO to reside in the most populated region of the Oph SFR. 
\cite{Evans.09.apjs} reported J163154.7-250324 to have an extinction-corrected temperature and a luminosity of 3100 K and 2.5$L_{\odot}$, respectively. It is listed as a YSO candidate star with extinction from dust, but no evidence of local extinction from a surrounding envelope. \cite{Cieza.09}, like \cite{Ratzka.05}, found no evidence for multiplicity. 

C10 identified the system as a TD and determined it to be a K7-type star, in agreement with our measurements above. Their \ion{Li}{1} measurements were too low in S/N and therefore unreported, so we cannot compare them to our own measurements of $0.5\pm0.2$\AA. They detected strong emission from the \ion{Ca}{2} triplet, providing further evidence that it is indeed a PMS object. 
Their measurement of the H$\alpha$ 10\% velocity-width (470 kms$^{-1}$) is similar to our own (364$\pm$61 kms$^{-1}$), but variation is expected due to variable accretion and the SB2 nature of the object. C10 found no evidence for a companion, as their study was not sensitive to separations $<\,8$ AU. They derive a limit on disk mass of $M_D\,<\,$1.1$\,M_{\rm Jup}$, and classify it as a grain growth-dominated TD (as mentioned in Section~\ref{intro}, this is their most common TD subclass). 

\section{Sample sensitivity to companions}
\label{complete}

We proceeded in a completeness study of our results following Section 6.1 of \cite{Duquennoy.91} and Section 3.1 of \cite{Melo03}, considering the observation cadence for each target, $v\sin i$, and the RV uncertainties of each spectrum. We sought to constrain how sensitive we were to SB1s within each TD for different orbital periods and secondary masses.

We estimated the primary mass of each target based on its absolute magnitude using low-mass solar metallicity stellar evolution models from \cite{Baraffe.15}. To estimate absolute magnitudes we apply the distance to Oph of 140\,pc, for Cha targets we used a distance of 160pc \citep{Feigelson.04}, and for CrA  and Cor targets we use a distance of 140\,pc \citep{S-A.08}. We use the \cite{Baraffe.15} 1 Myr CFHT tracks for all targets \citep[e.g.][]{Furlane.09, Kim.09}.
The 2MASS $J$ magnitudes, available for all of our targets (and $K$ where available), coupled with these age and distance estimates allowed us to estimate the mass of each target. We did not correct for extinction; NIR magnitudes are the least affected by optical- and NIR-excesses from accretion and inner disks. The average change in mass estimate between the \cite{Baraffe.98} and \cite{Baraffe.15} models was $|\Delta M|=0.2$\,M$_{\sun}$. Such a difference has little impact on our results.

With these estimated primary masses of each target, we generated $10^6$ artificial companions, each with mass ratio ($q$), period ($P$), time of periastron, inclination ($i$), and longitude of periastron ($\omega$) chosen randomly from a uniform distribution. Eccentricity ($e$) was chosen to be zero for $P <$ 8 days, and drawn from the Hyades distribution of \cite{Burki.86} for $P >$ 8 days. The solution of Kepler's Equation followed Chapter 2.5 of \cite{Hilditch.01}. At each date of observation, we calculated the RV for each artificial companion, and added to this a random uncertainty drawn from a Gaussian distribution centered around our measured uncertainty of the (real) primary RV (Table~\ref{table_rv}). 

To avoid an overestimate of detection probability at high mass ratios and long orbital periods (where we expect SB2 systems to lie in this ($M_{\rm secondary},P$)-space) we calculated the flux-weighted systemic velocity of the binary system

\begin{equation}
\gamma = \frac{{\rm RV}_1A_1 + {\rm RV}_2A_2}{A_1 + A_2}
\end{equation}

where $A_2$/$A_1$ is the flux ratio of the stars, and ${\rm RV}_1$ and ${\rm RV}_2$ are the RVs of each star. Their orbits are related by angle $\theta_2 = \theta_1 + \pi$, orbiting the center of mass. We interpolated the fluxes in the I-band using the \cite{Baraffe.15} 1 Myr CFHT tracks. Defining $\alpha_1 = \cos(\theta_1 + \omega) + e\cos\omega$, $\alpha_2 = \cos(\theta_1 + \pi + \omega) + e\cos\omega$, and expressing the flux ratio as $F_R = A_2/A_1$, the systemic RV of the system is then

\begin{equation}
\label{eq:Vavg}
\gamma = \frac{2\pi a \sin i}{(1+F_R)P(1-e^2)^{1/2}}(M_2\alpha_1 + M_1\alpha_2F_R)
\end{equation}

We then tested if the artificial companion could be `detected' by whether or not the root-mean-square of the values of $\gamma$ (each given by Equation~\ref{eq:Vavg}) was greater than 0.7 kms$^{-1}$, i.e. greater than the upper limit on the standard errors of our non-SB RVs. With over $10^6$ simulations per target, we were able to map probabilities of detection in ($M_{\rm secondary},P$)-space. The results of these simulations are shown in Figure~\ref{sblims}, which shows contour levels in $10$\% bins from $40-100$\% chance of detection\footnote{While contour levels run between these values, they rarely go below a 60\% chance of detection.  This happens only for the lowest secondary masses.}. The periodic modulation in the contour patterns demonstrates our insensitivity to orbital periods that are multiples of our observing cadence.

\begin{figure}
\centering
\includegraphics[width=0.95\textwidth]{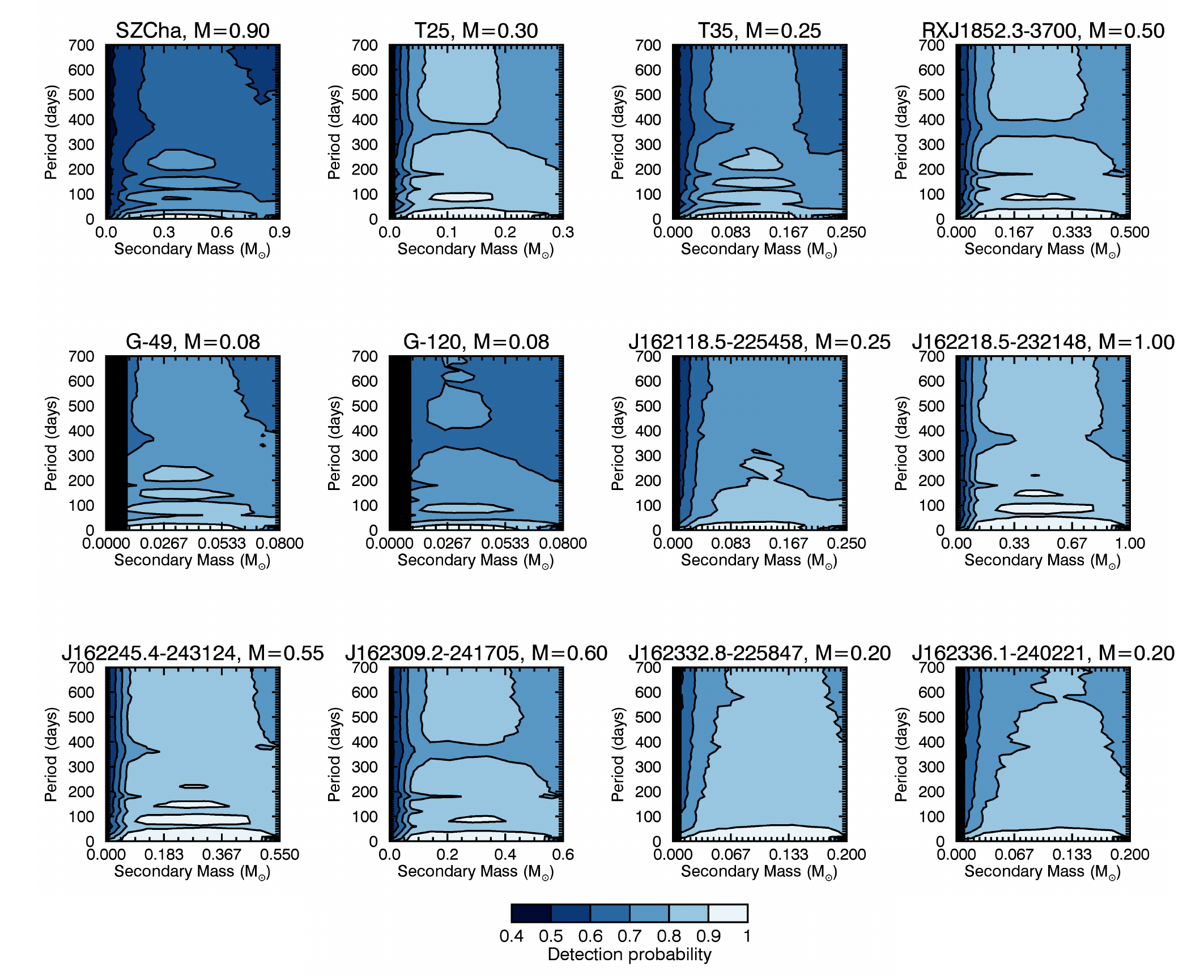}
\caption{Calculated contours in ($M_{\rm secondary},P$)-space representing the probability of detecting a companion star spectroscopically, given the precision and timing of our RV measurements. Contour levels of probability of detection are shown in $10$\% bins from $40-100$\%. 
Shown next to the name of the target system is the mass of the primary in solar masses, according to the PMS models of \protect\cite{Baraffe.15}. 
We imposed that the secondary mass not exceed the primary mass for each system. 
Two targets are not shown since we have only one RV measurement for the star, and therefore no chance of finding SB1s: these are SR21A and CrA-4111. For the Ophiuchus targets, we have dropped the `SSTc2d' prefix.}
\label{sblims}
\end{figure}
\addtocounter{figure}{-1}
\begin{figure}
\centering
\includegraphics[width=0.95\textwidth]{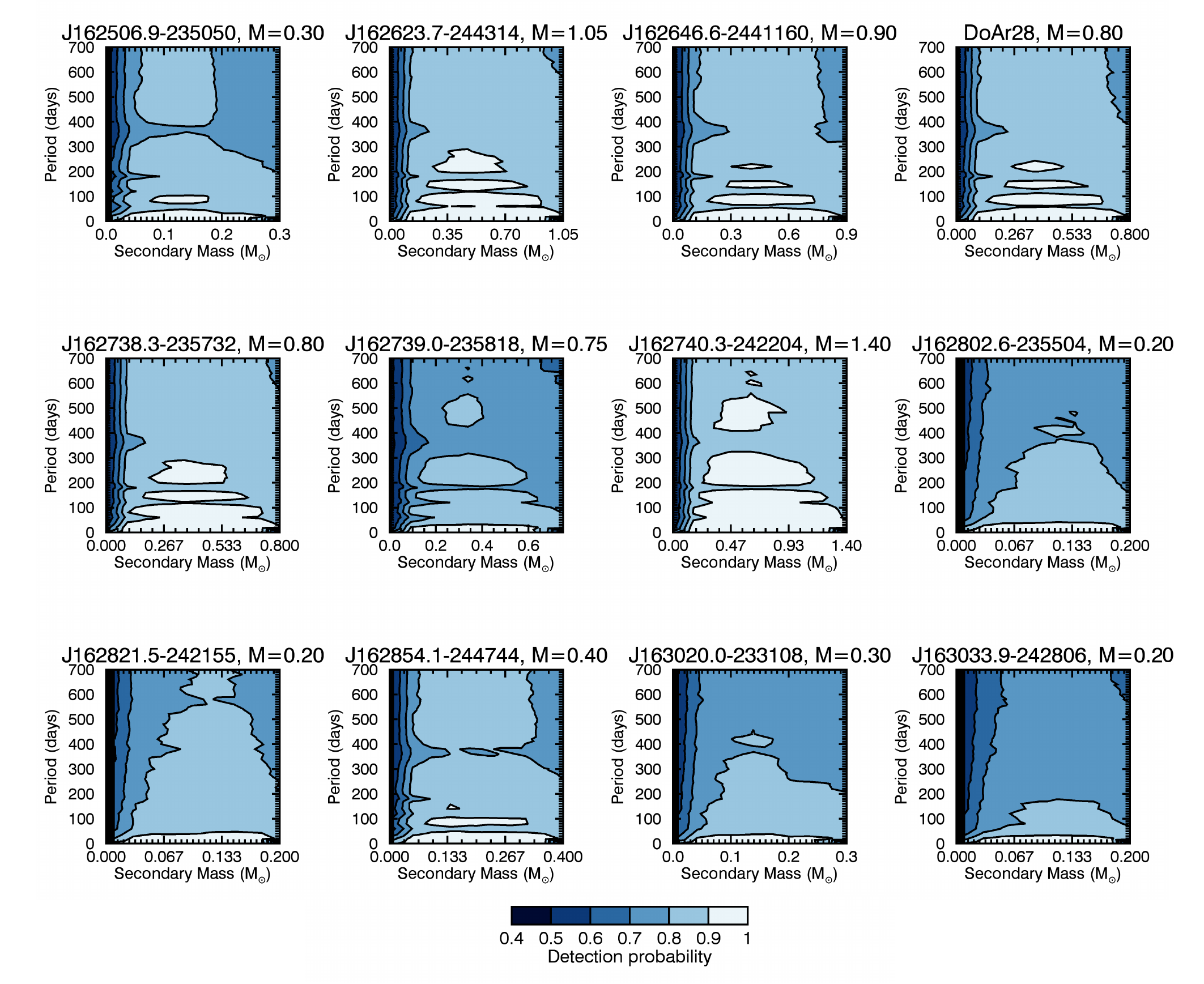}
\caption{(continued.)}
\end{figure}
\addtocounter{figure}{-1}
\begin{figure}[h]
\centering
\includegraphics[width=0.95\textwidth]{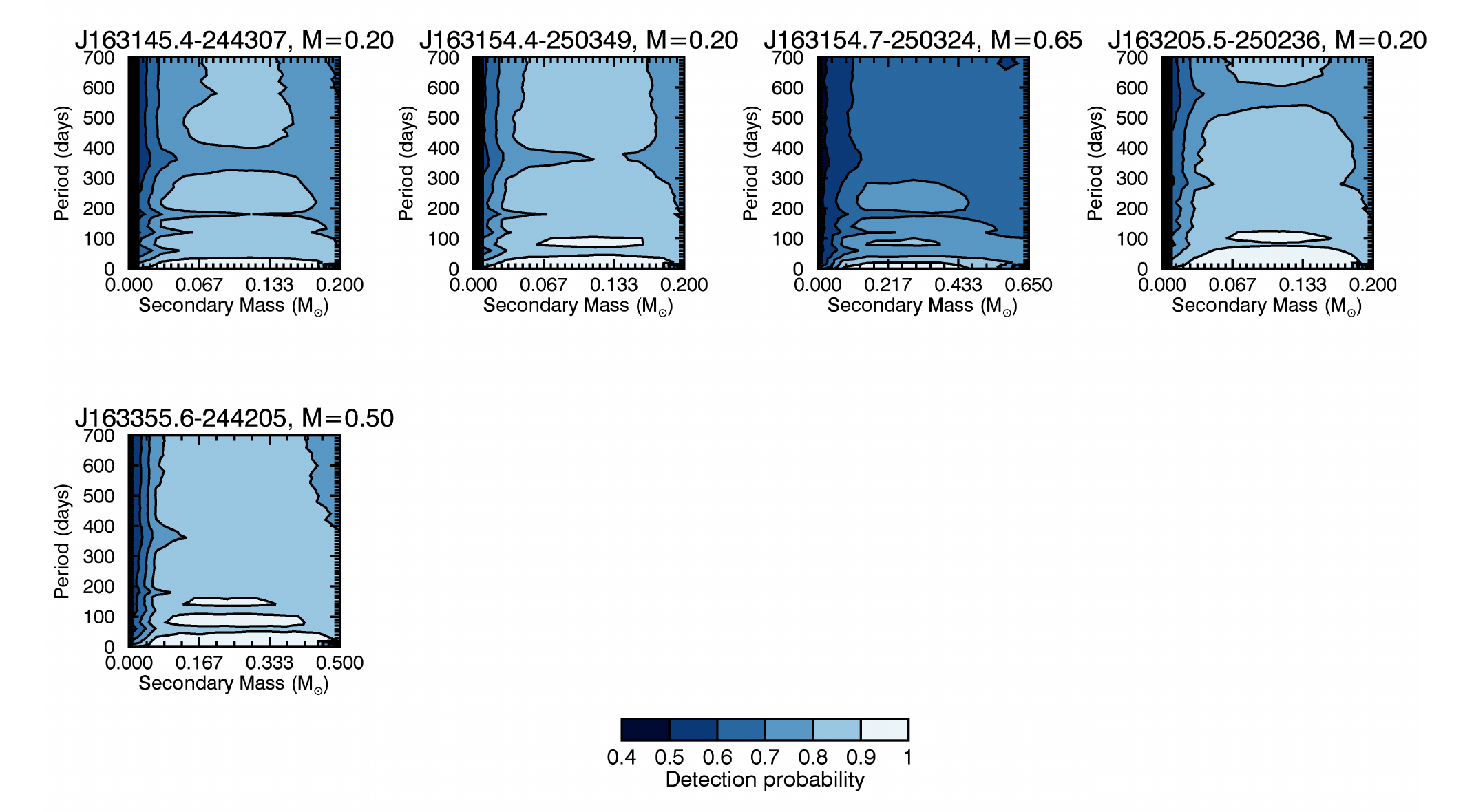}
\caption{(continued.)}
\end{figure}

The results of these simulations allow us to accurately assess our sensitivity to companions on a per-target basis, with a degeneracy in $P$ and mass ratio $q$.\footnote{One can convert between limits on the period and limits on the semi-major axis of separation $a$ using the relation: $a(P,q,M_{\rm primary}) = (P/2\pi)^{2/3}(G(1+q)M_{\rm primary})^{1/3}$}

For example, for SZ Cha we are sensitive to ($\geq$80\% chance of detection) a 0.4\,M$_{\sun}$ companion at periods $P\,<\,30$ days and  $70\,<\,P\,<\,75$ days, whereas for SSTc2d J162623.7-244314, we are sensitive to a 0.4\,M$_{\sun}$ companion at all periods. 
Such an example illustrates the challenge in quoting an ``overall" or ``average" sensitivity for this non-homogeneous sample. 
We are, however, consistently sensitive to short-period ($P\,<\,\sim100$ days) SBs for all of our targets.

\section{Discussion \label{disc}}

\cite{Prato07} measures a spectroscopic binary fraction of 0.12$^{+0.08}_{-0.04}$ among Oph K7-M4 stars. Of the 33 targets in \cite{Prato07}, two are TD objects but neither appears to be an SB: ROXR1 20 (C10) and RX	J1614.4-1857 \citep{Wahhaj.10}. Of the five SBs they found, one hosts a debris disk \citep[RX J1612.6-1924,][]{Wahhaj.10} and the rest are diskless (RX J1612.3-1909, RX J1622.7-2325, and RX J1622.8-2333; \citealt{Wahhaj.10}; ROXR1 14 \citealt{Cieza.07, Rosero.11}). We found one SB2 in a sample of 24 Oph TD objects, and no SBs among the TD objects in the other SFRs surveyed.
Including the two TD targets from \cite{Prato07} in our Oph sample, we find that multiplicity among TD objects in this region to be 1/26. We measure an SB fraction of 0.04$^{+0.12}_{-0.03}$, determining uncertainties following the binomial theorem.
This is consistent with that of the non-TD, late-type stars in Ophiuchus \citep{Prato07} and the young stars in Chameleon and Taurus-Auriga \citep[0.07$^{+0.05}_{-0.03}$ and 0.06$^{+0.03}_{-0.02}$, respectively; mass range $0.2$--$3\,M_{\odot}$;][]{Nguyen.09}. 
The result also in agreement with \cite{Lodieu.14}, who report an SB fraction of  $0.054 \pm 0.038$ for wide binaries ($50-150$ AU) in which one star is a planet-host. Their SB fraction for wide, planet-hosting substellar binaries is also within our range at $0.027 \pm 0.027$.
We also conclude that there is no significant difference between our result and that among the low-mass (0.08$M_{\odot}$--0.6$M_{\odot}$) field stars in the surveys of  \cite{Duquennoy.91} and \cite{Marcy92}, 0.09$^{+0.03}_{-0.02}$ and 0.03$^{+0.04}_{-0.02}$, respectively. 
Likewise, \cite{Raghavan.10} found an SB fraction of 0.073$^{+0.014}_{-0.012}$ among nearby old solar-type dwarfs.

That the fraction of SBs in TDs is similar to diskless stars (within uncertainties) may suggest that whatever causes the disk dissipation is independent of binarity, i.e. disks around close SBs evolve on a similar timescale as those around single stars. 
Such a conclusion would mean that planet formation timescales around SBs may be comparable to those around single stars.

However, several studies \citep{Ghez.97, White.01, Cieza.09, Duchene.10, Kraus.12} have found protoplanetary disk fractions are lower for young ($\leq 5$ Myr) binary systems with $a<50$ AU than for single stars or wider binaries. 
 With the exception of \cite{Kraus.12}, these studies were not sensitive to the tight binary systems to which our study was especially sensitive. 
\cite{Kraus.12} and \cite{Cheetham.15}, sensitive to similar binary separations to our study, find the disk fraction among close visual binaries ($0.1$--$40$\,AU) in Taurus-Auriga and Ophiuchus to be lower than for single stars. 
Their studies did not distinguish between circumbinary TDs and other disk types (e.g. debris disks).
If the SB fraction of TD systems is actually lower than that in the diskless population,  this suggests that the binary TD stage of a close-in binary is even shorter than the TD stage of single stars.

While we have searched for SBs specifically in TDs, \cite{Prato07} made a complementary search for SBs irrespective of knowledge of their disks. None of the SBs she found were in transitional or protoplanetary disks. This is consistent with SBs having a faster TD phase and that the overall disk fraction among SBs remains low down to small separations. 

Such a conclusion would be in contention with the results of \cite{Alexander.12}, whose models suggest that protoplanetary disks around tight binaries ($a\leq1$ AU) are longer-lived than those around wider binaries. This is because the very close binaries efficiently clear the inner disk to radii larger the characteristic radius of photoevaporative winds, suppressing total disk dispersal.

That we only find one SB in our TD sample could reflect that dissipation is so fast among these objects that most disks are completely dispersed by the age of Oph. However, we do not have the statistics to draw a steadfast conclusion.
But since we find no evidence that TD objects have higher SB fractions than other stellar populations, we can infer that close-in binaries are not likely the primary cause of inner holes in TDs. 
This leads to giant planet formation as a more likely explanation.


\section{Summary \label{conc}}

We have presented a spectroscopic survey of 31 TD stars, 24 of which lie in Ophiuchus. 
We found one of the Oph stars to be an SB2 (SSTc2d J163154.7-250324).
This system is composed of a K7($\pm$0.5) and a K9($\pm$0.5) star, with mass ratio $q\simeq$0.95 and orbital limits $a<$0.6 AU and $P<$150 days.
The average RV of our Oph targets is -6.6 kms$^{-1}$, with a dispersion of 1.3 kms$^{-1}$. The median uncertainty in the measured RVs including systematic uncertainties is 0.4 km s$^{-1}$. 
The average RVs of all four clouds surveyed (Cha, Cor, CrA and Oph) are consistent with the literature values.
With this single SB detection, we measure an SB fraction of 0.04$^{+0.12}_{-0.03}$. This finding is consistent with that of non-TD late-type stars in and outside of the region. This suggests that a TD may not be dispersed more efficiently by a tight binary than by a single star, and may imply that planet formation timescales around close binaries are comparable to those around single stars. 

\section*{Acknowledgements}
We thank M. Hughes for her contribution of RX J1852.3-3700 to the target list prior to its publication; L.~Prato, C.~Johns-Krull and C.~H.~Blake for helpful discussions; and the anonymous referee for her/his insightful comments. S.~A.~Kohn acknowledges the support of NSF REU grant AST-1004107 through Northern Arizona University and Lowell Observatory. J. Llama acknowledges support from NASA Origins of the Solar System grant No. NNX13AH79G and from STFC grant ST/M001296/1. This research has made use of the SIMBAD database, operated at CDS, Strasbourg, France.\\ 
\textit{Facility}: LCO: Magellan (MIKE).

\section{Appendix}
\label{App}

We present here the accretion properties and \ha and \ion{Li}{1} measurements of individual stars.

\subsection{Age and accretion diagnostics} 
\label{age_accretion}
The strength of the \ion{Li}{1} line ($\lambda$=6708 \AA) is a standard diagnostic for the age of low-mass stars, since lithium is rapidly depleted in their atmospheres. All but one of our targets exhibit \ion{Li}{1} absorption, so we are only able to place upper limits on the absorption in the spectra of CrA-4111 due to low S/N in our single observation of the system.
The \ion{Li}{1} should be undetectable in early M stars after $\sim$20 Myr \citep{Baraffe.98, White.05}. For G and K stars the depletion can take much longer \citep[$\geq 300$ Myr, e.g.][]{Zickgraf.05}.
The age of Oph is $<$10 Myr \citep[][C10]{Duncan.81, Chabrier96, Martin.99, Martin.99b, Zuckerman.04}, and thus, all stars are expected to display Li absorption. 
For the 12 stars at sufficient S/N to detect Li that were also observed in C10, our EWs are consistent with theirs. Our \ion{Li}{1} EW measurements of DoAr 28 and SR21A are consistent with those of \cite{Magazzu.92} and \cite{James.06}, respectively. We are not aware of any previous measurements of \ion{Li}{1} in SSTc2d J162309.2-241705. The  \ion{Li}{1} EWs are listed in Table~\ref{table_rv}.

\begin{figure}[ht!]
\caption{Variable accretion shown in \ha.}
\label{fig_var_acc}
\begin{center}
\begin{tabular}{cc}
\includegraphics[scale=0.3]{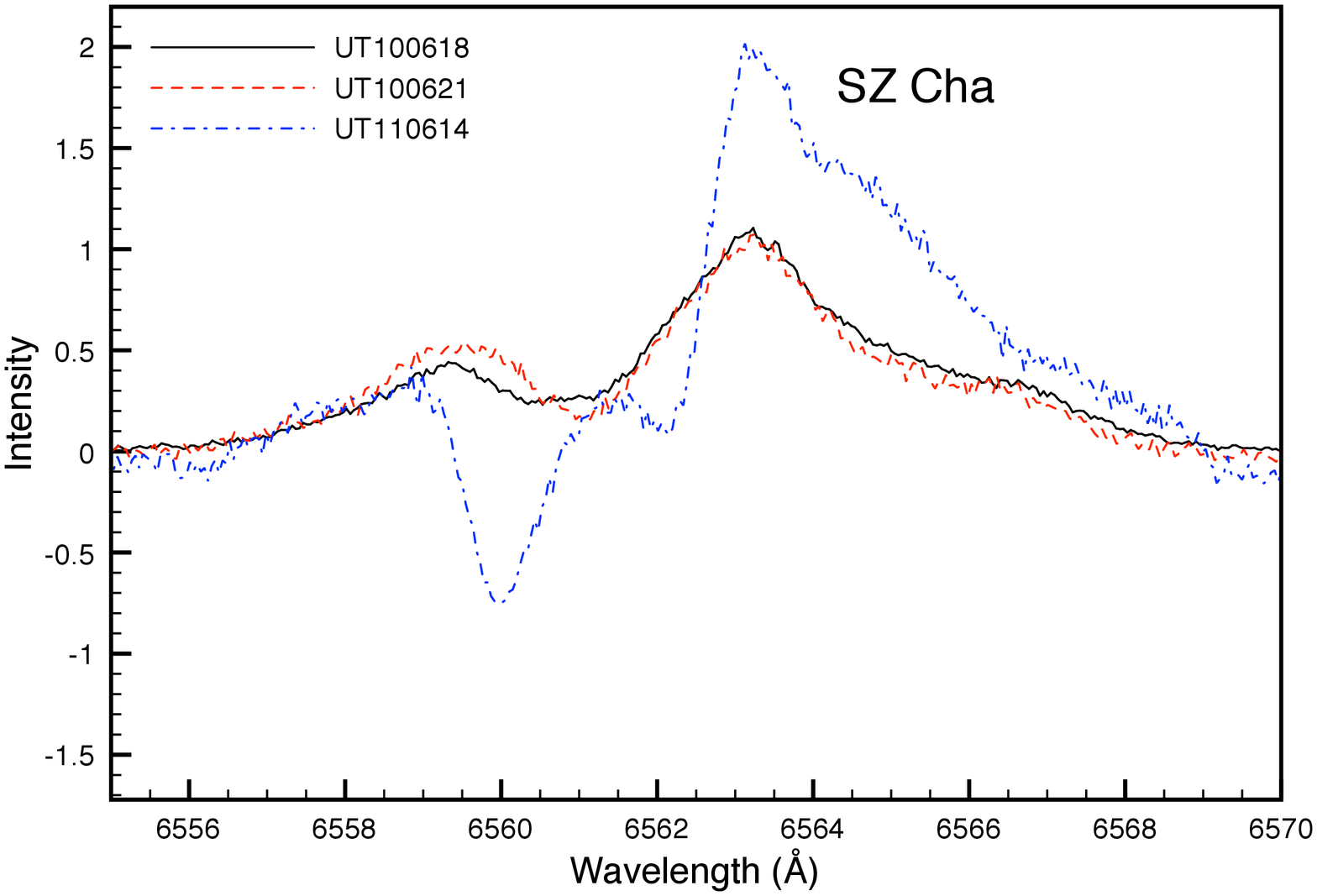}&
\includegraphics[scale=0.3]{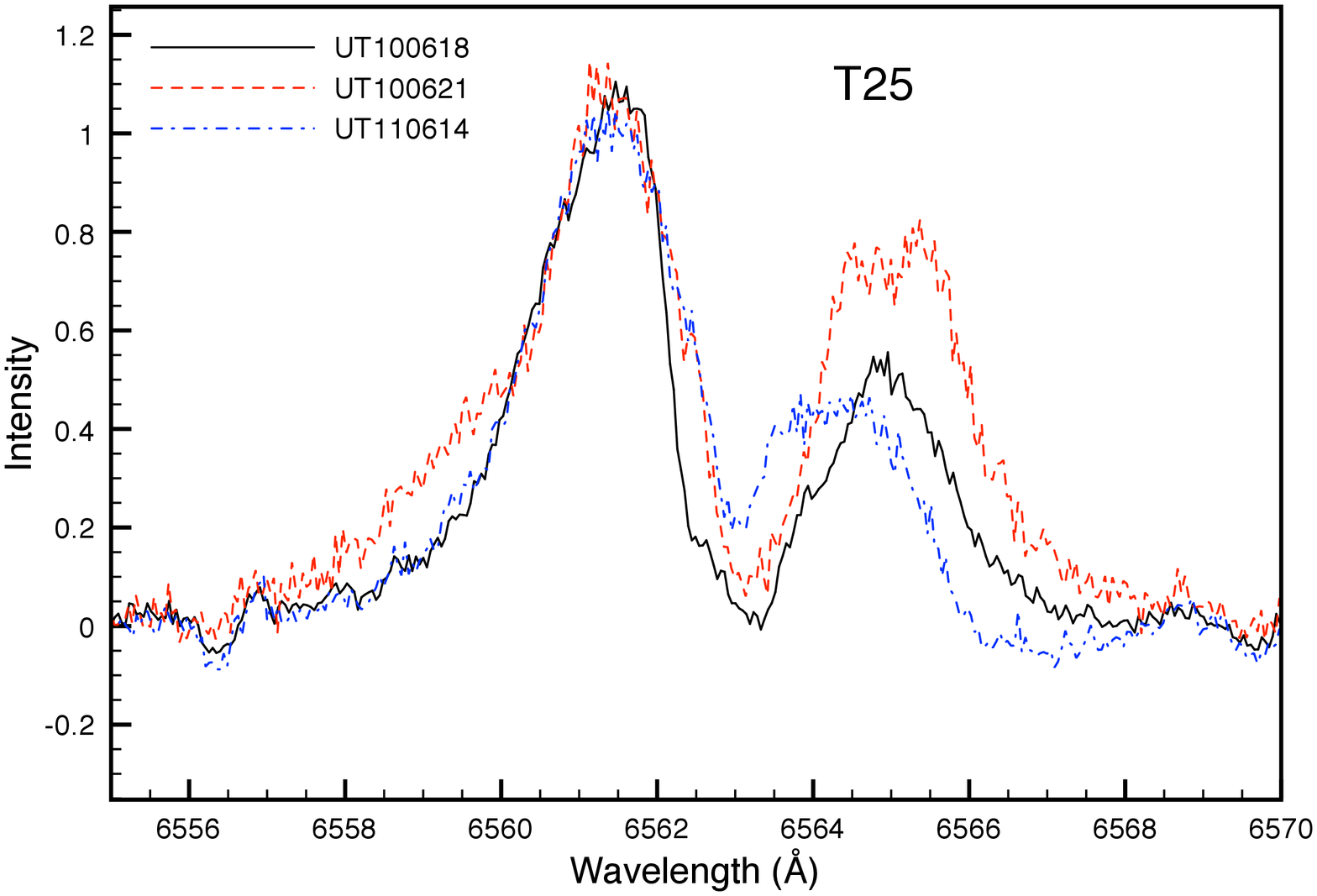}\\
\includegraphics[scale=0.3]{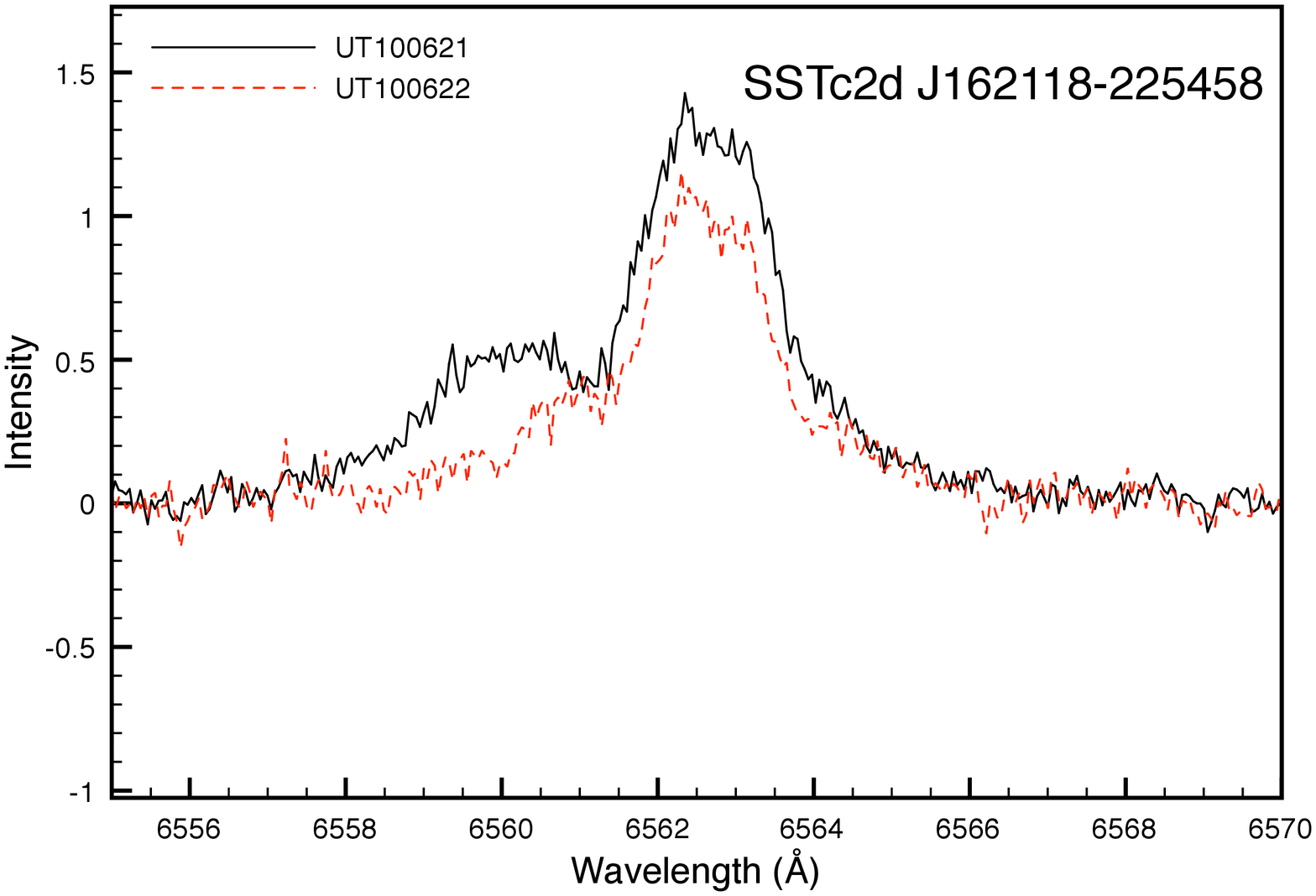}&
\includegraphics[scale=0.3]{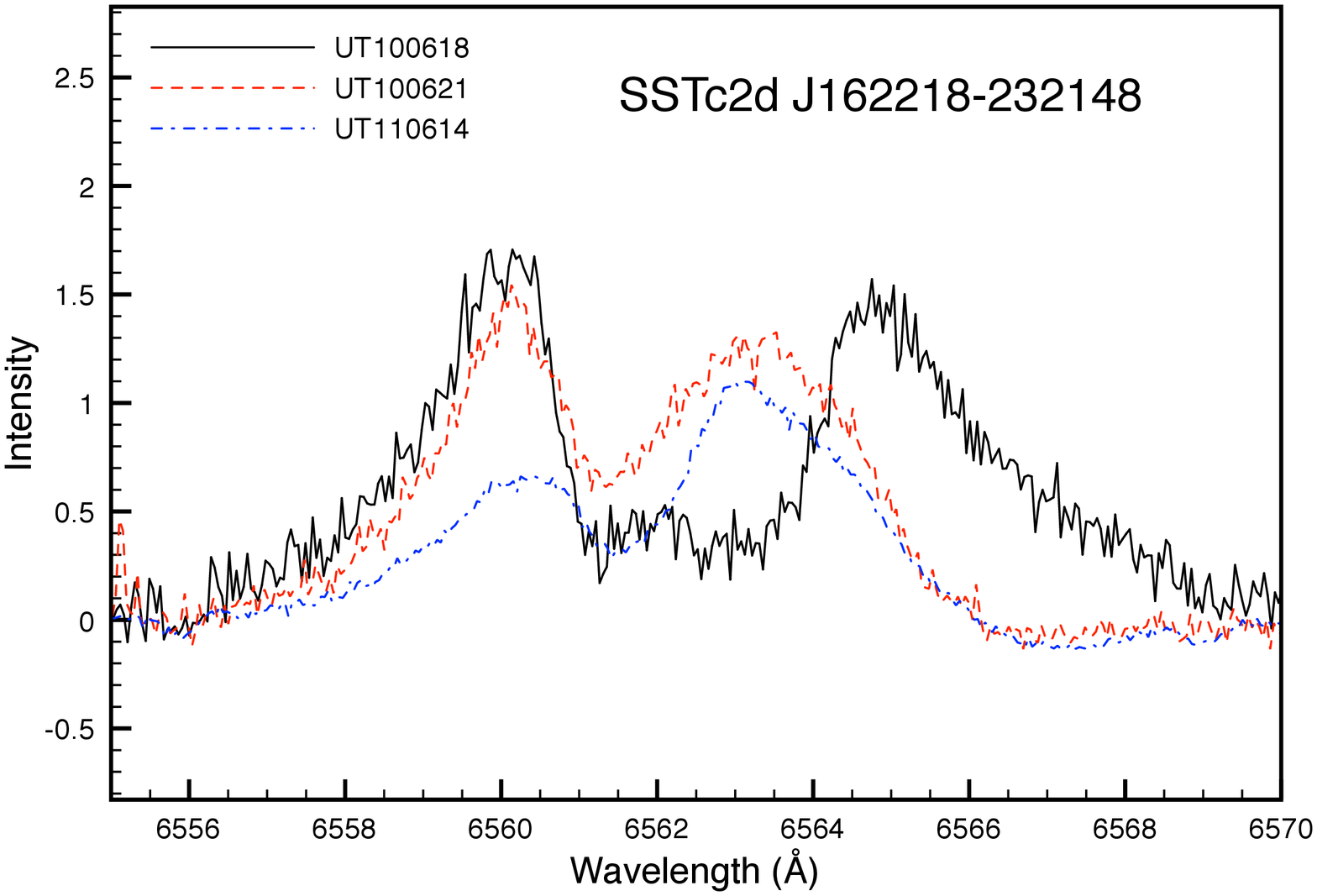}\\
\includegraphics[scale=0.3]{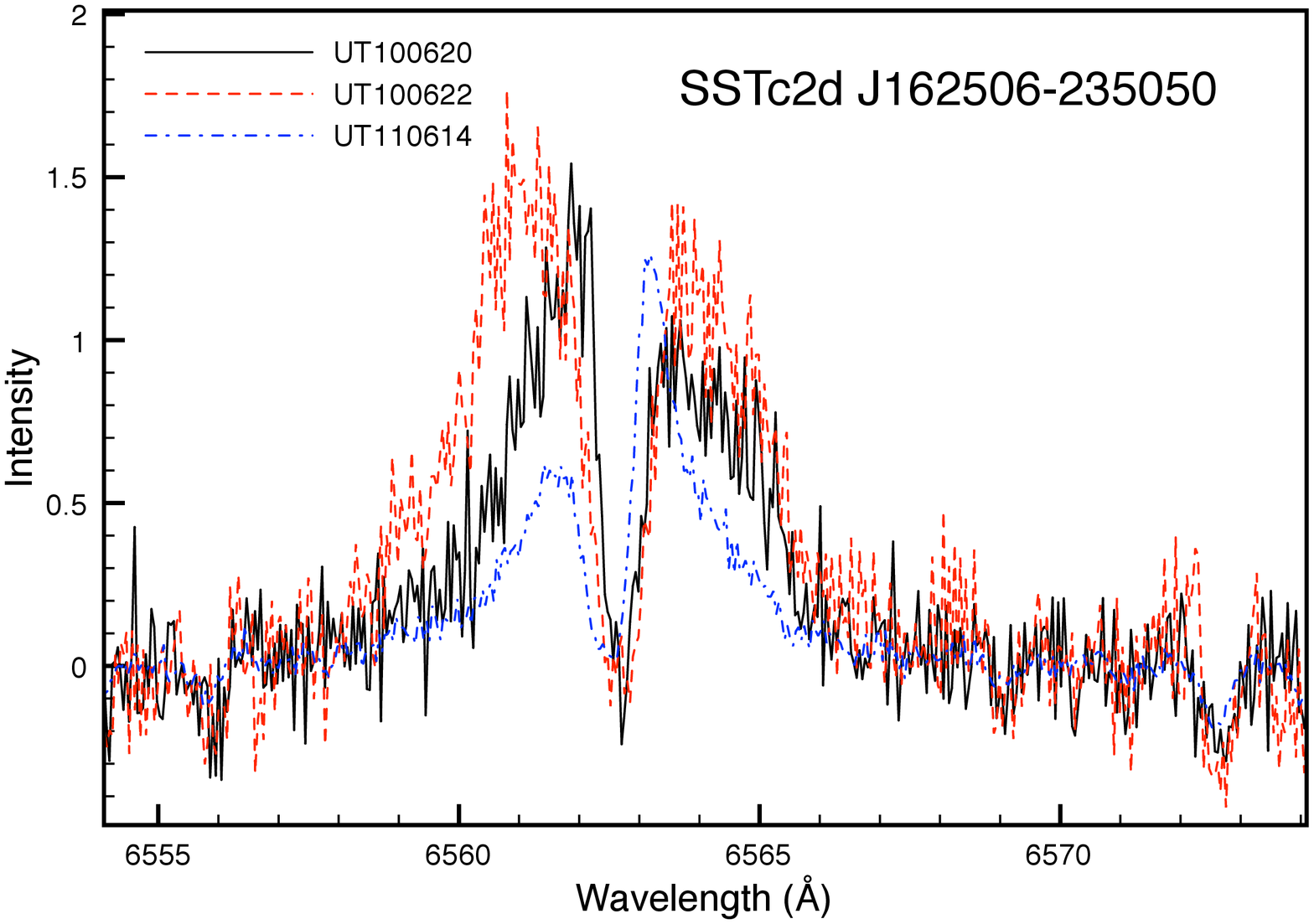}&
\includegraphics[scale=0.3]{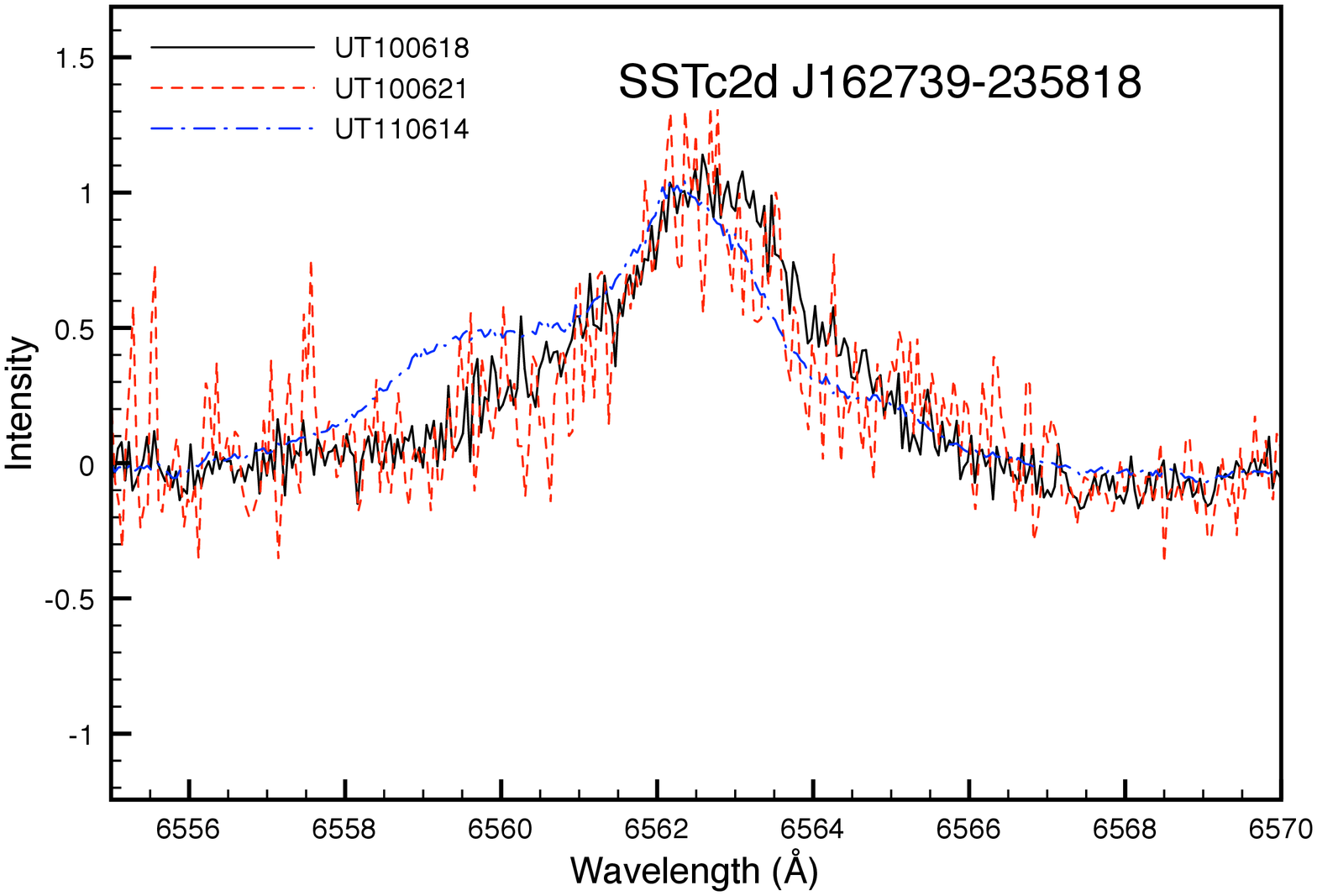}\\
\end{tabular}
\end{center}
\end{figure}

\ha ($\lambda$=6563 \AA) is an indicator of stellar activity and gas accretion onto a star. Applying the relation found by \cite{Natta.04}, we use the 10\% velocity-width \citep{WhiteBasri.03} of the H$\alpha$ emission  to estimate the rate of accretion from the disk onto the star for targets with H$\alpha$ 10\% velocity-widths greater than 200 kmbs$^{-1}$ (this is true for 21 of our 31 targets). We find that of the 15 targets that were observed by C10 and had sufficient S/N at H$\alpha$, 10 have 10\% velocity-widths within 3$\sigma$ of those measured by C10. The large variations in the remaining measurements are likely due to variable accretion upon the star \citep[e.g.][]{Nguyen.09}. 
Six of our targets exhibited varying \ha\ during our observations as shown in Figure~\ref{fig_var_acc}. 
These are SZ Cha, T25, SSTc2d J162118-225458, SSTc2d J162218-232148, SSTc2d J162739-235818, and SSTc2d J162506-235050. The level of variability is unique for each target, but on average the EW varies by $\sim$80\%.

\subsection{SR21A}
\label{sr21a}
\begin{figure}[ht!]
\centering
\begin{tabular}{cc}
\includegraphics[scale=0.3]{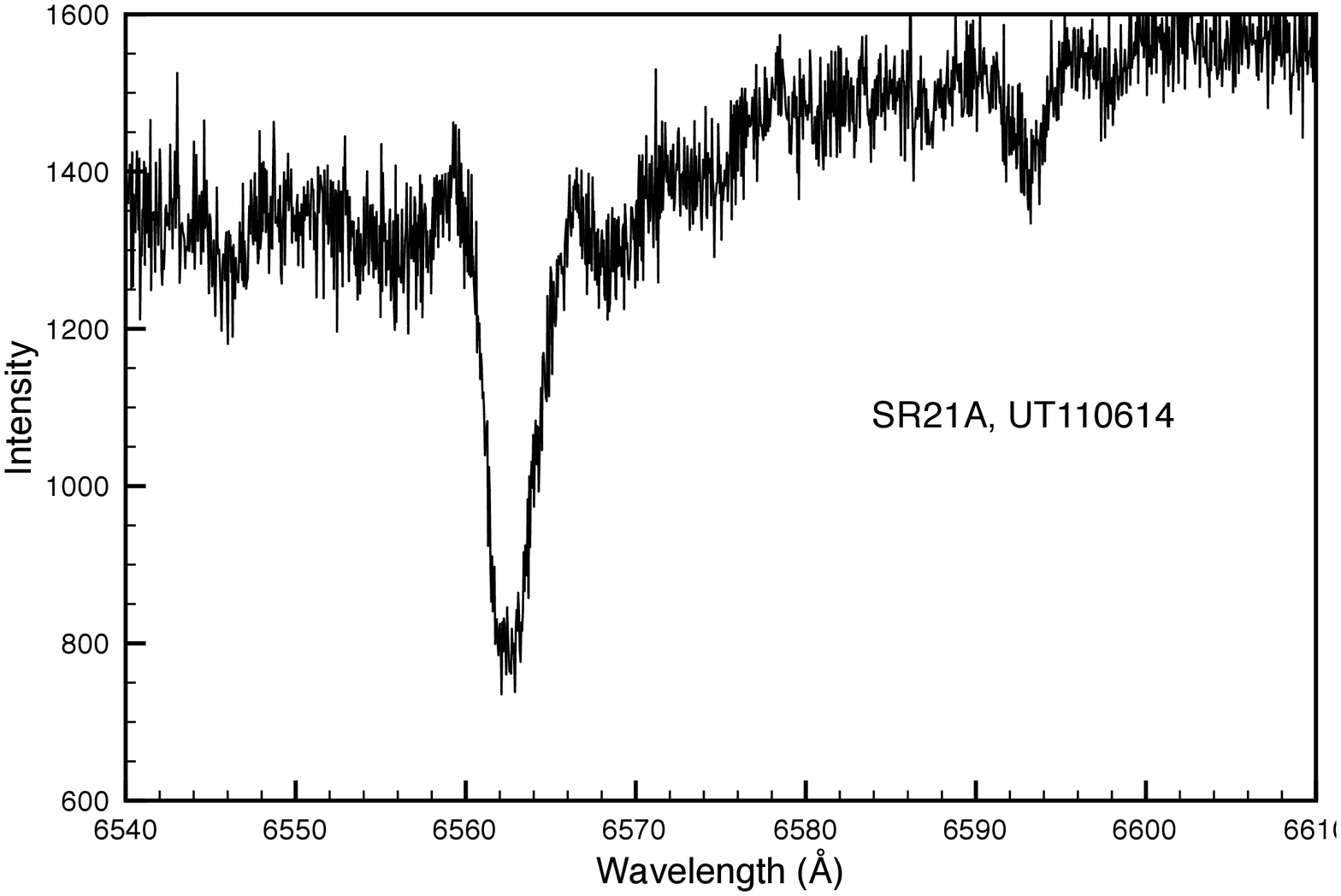} & \includegraphics[scale=0.3]{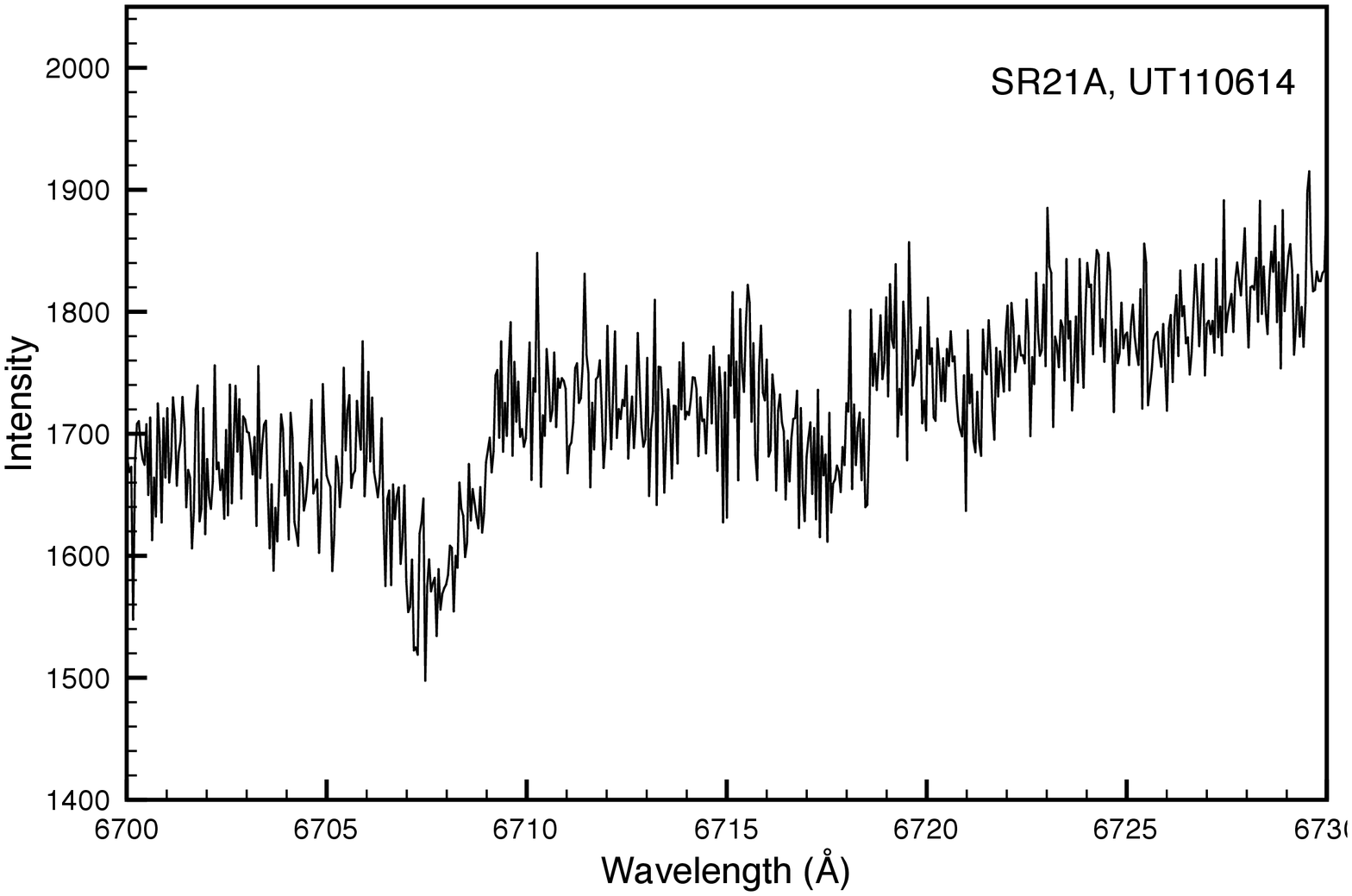}
\end{tabular}
\caption{Spectral features of SR21A: \ha absorption (\textit{left}; EW=1.13$\pm$0.15 \AA) and weak \ion{Li}{1} absorption (\textit{right}; EW=0.13$\pm$0.02 \AA).}
\label{fig_SR21A}    
\end{figure}

At about 1--3 Myr old \citep[which we note is older than the average age of Oph stars;][]{Seiss.00, Andrews.09, Brown.09, Follette.13}, SR21A is the primary component of a wide binary system in Oph \citep{Barsony.05}. Its spectral type has been reported to be an early G  \citep{Suarez.06, Andrews.07}, an early K \citep{Rudkjobing49}, and an M star \citep{Kimeswenger.04}. 
Our measurements of its TiO-7140 index (see Section~\ref{subsec:calcSpTs}) suggest it is not an M-type star (SpT$_{\rm TiO}\,<0$).
SR21A displays \ha in absorption rather than in emission,\footnote{It should be noted that SR21A is listed as an emission star in the SIMBAD database, but we did not observe is as such. \cite{Rudkjobing49} do list it as a non-emitting star in their spectral emission study of Oph.} the only such star in our sample. \ion{Li}{1} absorption is also weaker than all of our other Oph objects. 
The \ha absorption could be due to variable accretion \citep{Johns.01, Baraffe.12, Hamilton.12}. This is suggested by the broadened blue shoulder of the absorption (a feature possibly hidden by the \ion{N}{2} absorption on the red shoulder), a sign of possible accretion emission superimposing upon the photosphere absorption (Figure~\ref{fig_SR21A}, left). 
Weak \ion{Li}{1} absorption (0.13$\pm$0.02 \AA; Figure~\ref{fig_SR21A}, right) might suggest that SR21A is older than expected ($>$10Myr; \citealt{Chabrier96}), but the existence of a transition disk is evidence against the star being $>$15Myr \citep{Mamajek.09, Muzerolle.10}. 
While \cite{James.06} find a slightly higher EW for \ion{Li}{1} in SR21A ($0.275 \pm 0.028$ \AA , versus our value of 0.13$\pm$0.02 \AA), they also note its inconsistency with other EWs in Oph. 
It is possible that the same episodic accretion that could be causing the \ha absorption has also resulted in abnormal Li depletion in this star \citep{Baraffe.09, Baraffe.10, Baraffe.12}. 
Depletion is also more realistic than the feature being veiled by accretion, since a disk hot enough for veiling would have caused SR21A to be disqualified as a TD object. 
However, we compared the EWs/line strengths of all of the lines in the same echelle order as the Li line in SR21A with the late K- and early M-type stars we observed on the same night (UT110614). We found that SR21A had weaker absorption for all lines, suggesting that veiling may still play a role. Overall weaker lines might instead indicate a lower overall metallicity compared with the other Oph targets.

An extensive study of the disk morphology of SR21A was recently presented by \cite{Follette.13}. Their results may support the postulate of a substellar companion present within the SR21A disk at $\sim$18 AU from the primary \citep{Eisner.09}. Since we have just one observation of this system, we can neither confirm nor exclude this finding.

\clearpage
\bibliographystyle{apj}
\bibliography{myBib.bib}{}

\begin{thebibliography}{138}
\expandafter\ifx\csname natexlab\endcsname\relax\def\natexlab#1{#1}\fi

\bibitem[{{Alexander}(2012)}]{Alexander.12}
{Alexander}, R. 2012, ApJL, 757, L29

\bibitem[{{Allers} {et~al.}(2006){Allers}, {Kessler-Silacci}, {Cieza}, \&
  {Jaffe}}]{Allers.06}
{Allers}, K.~N., {Kessler-Silacci}, J.~E., {Cieza}, L.~A., \& {Jaffe}, D.~T.
  2006, ApJ, 644, 364

\bibitem[{Andrews \& Williams(2007{\natexlab{a}})}]{AndrewsWilliams.07}
Andrews, S., \& Williams, J. 2007{\natexlab{a}}, ApJ, 671, 1800

\bibitem[{Andrews \& Williams(2007{\natexlab{b}})}]{Andrews.07}
Andrews, S.~M., \& Williams, J.~P. 2007{\natexlab{b}}, ApJ, 671, 1800

\bibitem[{Andrews {et~al.}(2011)Andrews, Wilner, Espaillat, Dullemond, McClure,
  \& Qi}]{Andrews.11}
Andrews, S.~M., Wilner, D.~J., Espaillat, Catherine;~Hughes, A.~M., Dullemond,
  C.~P., McClure, M.~K., \& Qi, Chunhua;~Brown, J.~M. 2011, ApJ, 732, 25pp

\bibitem[{Andrews {et~al.}(2009)Andrews, Wilner, Hughes, \& Qi}]{Andrews.09}
Andrews, S.~M., Wilner, D.~J., Hughes, A.~M., \& Qi, Chunhua;~Dullemond, C.~P.
  2009, ApJ, 700, 1502

\bibitem[{Artymowicz \& Lubow(1994)}]{Lubow94}
Artymowicz, \& Lubow. 1994, ApJ, 421, 651

\bibitem[{Baraffe \& Chabrier(2010)}]{Baraffe.10}
Baraffe, I., \& Chabrier, G. 2010, A\&A, 521

\bibitem[{{Baraffe} {et~al.}(1998){Baraffe}, {Chabrier}, {Allard}, \&
  {Hauschildt}}]{Baraffe.98}
{Baraffe}, I., {Chabrier}, G., {Allard}, F., \& {Hauschildt}, P.~H. 1998, A\&A,
  337, 403

\bibitem[{Baraffe {et~al.}(2009)Baraffe, Chabrier, \& Gallardo}]{Baraffe.09}
Baraffe, I., Chabrier, G., \& Gallardo, J. 2009, ApJL, 702, L27

\bibitem[{{Baraffe} {et~al.}(2015){Baraffe}, {Homeier}, {Allard}, \&
  {Chabrier}}]{Baraffe.15}
{Baraffe}, I., {Homeier}, D., {Allard}, F., \& {Chabrier}, G. 2015, A\&A, 577,
  A42

\bibitem[{Baraffe {et~al.}(2012)Baraffe, Vorobyov, \& Chabrier}]{Baraffe.12}
Baraffe, I., Vorobyov, E., \& Chabrier, G. 2012, ApJ, 756, 118

\bibitem[{Barsony \& Marsh(2005)}]{Barsony.05}
Barsony, Mary;~Ressler, M.~E., \& Marsh, K.~A. 2005, ApJ, 630, 381

\bibitem[{{Beichman} {et~al.}(1988){Beichman}, {Neugebauer}, {Habing}, {Clegg},
  \& {Chester}}]{Beichman.88}
{Beichman}, C.~A., {Neugebauer}, G., {Habing}, H.~J., {Clegg}, P.~E., \&
  {Chester}, T.~J., eds. 1988, {Infrared astronomical satellite (IRAS) catalogs
  and atlases. Volume 1: Explanatory supplement}, Vol.~1

\bibitem[{Berdyugina(2005)}]{Berdyugina.05}
Berdyugina, S.~V. 2005, Living Rev. Solar Phys. 2

\bibitem[{{Bernstein} {et~al.}(2003){Bernstein}, {Shectman}, {Gunnels},
  {Mochnacki}, \& {Athey}}]{Bernstein.03}
{Bernstein}, R., {Shectman}, S.~A., {Gunnels}, S.~M., {Mochnacki}, S., \&
  {Athey}, A.~E. 2003, in Society of Photo-Optical Instrumentation Engineers
  (SPIE) Conference Series, Vol. 4841, Society of Photo-Optical Instrumentation
  Engineers (SPIE) Conference Series, ed. M.~{Iye} \& A.~F.~M. {Moorwood},
  1694--1704

\bibitem[{{Beuther} {et~al.}(2014){Beuther}, {Klessen}, {Dullemond}, \&
  {Henning}}]{Beuther.14}
{Beuther}, H., {Klessen}, R., {Dullemond}, C., \& {Henning}, T. 2014,
  Protostars and Planets VI, Space Science Series (University of Arizona Press)

\bibitem[{{Biller} {et~al.}(2012){Biller}, {Lacour}, {Juh{\'a}sz}, {Benisty},
  {Chauvin}, {Olofsson}, {Pott}, {M{\"u}ller}, {Sicilia-Aguilar}, {Bonnefoy},
  {Tuthill}, {Thebault}, {Henning}, \& {Crida}}]{Biller.12}
{Biller}, B., {et~al.} 2012, ApJL, 753, L38

\bibitem[{Bontemps {et~al.}(2001)Bontemps, Andre, Kaas, Nordh, Olofsson,
  Huldtgren, Abergel, Blommaert, Boulanger, Burgdorf, Cesarsky, Cesarsky,
  Copet, Davies, Falgarone, Lagache, Montmerle, Perault, Persi, Prusti, Puget,
  \& Sibille}]{Bontemps.01}
Bontemps, S., {et~al.} 2001, A\&A, 372, 173

\bibitem[{{Brown} {et~al.}(2009){Brown}, {Blake}, {Qi}, {Dullemond}, {Wilner},
  \& {Williams}}]{Brown.09}
{Brown}, J.~M., {Blake}, G.~A., {Qi}, C., {Dullemond}, C.~P., {Wilner}, D.~J.,
  \& {Williams}, J.~P. 2009, ApJ, 704, 496

\bibitem[{{Burki} \& {Mayor}(1986)}]{Burki.86}
{Burki}, G., \& {Mayor}, M. 1986, in IAU Symposium, Vol. 118, Instrumentation
  and Research Programmes for Small Telescopes, ed. J.~B. {Hearnshaw} \& P.~L.
  {Cottrell}, 385--399

\bibitem[{{Carroll}(1933)}]{Carroll.33}
{Carroll}, J.~A. 1933, MNRAS, 93, 478

\bibitem[{{Casagrande} {et~al.}(2008){Casagrande}, {Flynn}, \&
  {Bessell}}]{Casagrande.08}
{Casagrande}, L., {Flynn}, C., \& {Bessell}, M. 2008, MNRAS, 389, 585

\bibitem[{Chabrier {et~al.}(1996)Chabrier, Baraffe, \& Plez}]{Chabrier96}
Chabrier, G., Baraffe, I., \& Plez, B. 1996, ApJ, 459, L91

\bibitem[{{Cheetham} {et~al.}(2015){Cheetham}, {Kraus}, {Ireland}, {Cieza},
  {Rizzuto}, \& {Tuthill}}]{Cheetham.15}
{Cheetham}, A.~C., {Kraus}, A.~L., {Ireland}, M.~J., {Cieza}, L., {Rizzuto},
  A.~C., \& {Tuthill}, P.~G. 2015, ApJ, 813, 83

\bibitem[{Cieza {et~al.}(2007)Cieza, Padgett, Stapelfeldt, Augereau, Harvey,
  Evans, Merin, Koerner, Sargent, van Dishoeck, Allen, Blake, Brooke, Chapman,
  Huard, Lai, Mundy, Myers, Spiesman, \& Wahhaj}]{Cieza.07}
Cieza, L., {et~al.} 2007, ApJ, 667, 308

\bibitem[{Cieza {et~al.}(2009)Cieza, Padgett, Allen, McCabe, Brooke, Carey,
  Chapman, Fukagawa, Noriga-Crespo, \& Rebull}]{Cieza.09}
Cieza, L.~A., {et~al.} 2009, ApJ, 696, L84

\bibitem[{Cieza {et~al.}(2010)Cieza, Schreiber, Romero, Mora, Merin, Orellana,
  Harvey, \& Evans}]{Cieza.10}
Cieza, L.~A., Schreiber, M.~R., Romero, G.~A., Mora, M.~D., Merin,
  Bruno;~Swift, J.~J., Orellana, Mariana;~Williams, J.~P., Harvey, P.~M., \&
  Evans, Neal~J., I. 2010, ApJ, 712, 925

\bibitem[{{Claret}(2000)}]{Claret.00}
{Claret}, A. 2000, A\&A, 363, 1081

\bibitem[{{Cruz} \& {Reid}(2002)}]{Cruz.02}
{Cruz}, K.~L., \& {Reid}, I.~N. 2002, AJ, 123, 2828

\bibitem[{{Cutri} {et~al.}(2003){Cutri}, {Skrutskie}, {van Dyk}, {Beichman},
  {Carpenter}, {Chester}, {Cambresy}, {Evans}, {Fowler}, {Gizis}, {Howard},
  {Huchra}, {Jarrett}, {Kopan}, {Kirkpatrick}, {Light}, {Marsh}, {McCallon},
  {Schneider}, {Stiening}, {Sykes}, {Weinberg}, {Wheaton}, {Wheelock}, \&
  {Zacarias}}]{Cutri.03}
{Cutri}, R.~M., {et~al.} 2003, {2MASS All Sky Catalog of point sources.}

\bibitem[{Daemgen {et~al.}(2007)Daemgen, Siegler, Reid, \& Close}]{Daemgen.07}
Daemgen, S., Siegler, N., Reid, I.~N., \& Close, L.~M. 2007, AJ, 654, 558

\bibitem[{D'Alessio {et~al.}(2005)D'Alessio, Merin, Calvet, Hartmann, \&
  Montesinos}]{D.Alessio.05}
D'Alessio, P., Merin, B., Calvet, N., Hartmann, L., \& Montesinos, B. 2005,
  Revista Mexicana de Astronom{\'\i}a y Astrof{\'\i}sica, 41, 61

\bibitem[{{Desort} {et~al.}(2007){Desort}, {Lagrange}, {Galland}, {Udry}, \&
  {Mayor}}]{Desort.07}
{Desort}, M., {Lagrange}, A.-M., {Galland}, F., {Udry}, S., \& {Mayor}, M.
  2007, A\&A, 473, 983

\bibitem[{Dodson-Robinson \& Salyk(2011)}]{Dodson-Robinson.11}
Dodson-Robinson, S.~E., \& Salyk, C. 2011, ApJ, 738, 131

\bibitem[{{Doyle} {et~al.}(2011){Doyle}, {Carter}, {Fabrycky}, {Slawson},
  {Howell}, {Winn}, {Orosz}, {Pr{\'c}aronsa}, {Welsh}, {Quinn}, {Latham},
  {Torres}, {Buchhave}, {Marcy}, {Fortney}, {Shporer}, {Ford}, {Lissauer},
  {Ragozzine}, {Rucker}, {Batalha}, {Jenkins}, {Borucki}, {Koch}, {Middour},
  {Hall}, {McCauliff}, {Fanelli}, {Quintana}, {Holman}, {Caldwell}, {Still},
  {Stefanik}, {Brown}, {Esquerdo}, {Tang}, {Furesz}, {Geary}, {Berlind},
  {Calkins}, {Short}, {Steffen}, {Sasselov}, {Dunham}, {Cochran}, {Boss},
  {Haas}, {Buzasi}, \& {Fischer}}]{Doyle.11}
{Doyle}, L.~R., {et~al.} 2011, Science, 333, 1602

\bibitem[{{Ducati}(2002)}]{Ducati.02}
{Ducati}, J.~R. 2002, VizieR Online Data Catalog, 2237, 0

\bibitem[{{Duch{\^e}ne}(2010)}]{Duchene.10}
{Duch{\^e}ne}, G. 2010, ApJL, 709, L114

\bibitem[{{Duncan}(1981)}]{Duncan.81}
{Duncan}, D.~K. 1981, ApJ, 248, 651

\bibitem[{{Duquennoy} \& {Mayor}(1991)}]{Duquennoy.91}
{Duquennoy}, A., \& {Mayor}, M. 1991, A\&A, 248, 485

\bibitem[{{Eisner} {et~al.}(2009){Eisner}, {Monnier}, {Tuthill}, \&
  {Lacour}}]{Eisner.09}
{Eisner}, J.~A., {Monnier}, J.~D., {Tuthill}, P., \& {Lacour}, S. 2009, ApJL,
  698, L169

\bibitem[{Erickson {et~al.}(2011)Erickson, Wilking, Meyer, Robinson, \&
  Stephenson}]{Erickson.11}
Erickson, K.~L., Wilking, B.~A., Meyer, M.~R., Robinson, J.~G., \& Stephenson,
  L.~N. 2011, AJ, 142, id.140

\bibitem[{{Espaillat} {et~al.}(2012){Espaillat}, {Ingleby}, {Hern{\'a}ndez},
  {Furlan}, {D'Alessio}, {Calvet}, {Andrews}, {Muzerolle}, {Qi}, \&
  {Wilner}}]{Espaillat.12}
{Espaillat}, C., {et~al.} 2012, ApJ, 747, 103

\bibitem[{Evans {et~al.}(2003)Evans, Allen, Blake, Boogert, Bourke, Kessler,
  Koerner, Lee, Myers, Padgett, Pontoppidan, Sargent, Stapelfeldt, van
  Dishoeck, Young, \& Young}]{Evans.03}
Evans, Neal~J., I., {et~al.} 2003, PASP, 115, 965

\bibitem[{{Evans} {et~al.}(2009){Evans}, {Dunham}, {J{\o}rgensen}, {Enoch},
  {Mer{\'{\i}}n}, {van Dishoeck}, {Alcal{\'a}}, {Myers}, {Stapelfeldt},
  {Huard}, {Allen}, {Harvey}, {van Kempen}, {Blake}, {Koerner}, {Mundy},
  {Padgett}, \& {Sargent}}]{Evans.09.apjs}
{Evans}, II, N.~J., {et~al.} 2009, ApJS, 181, 321

\bibitem[{{Feigelson} \& {Lawson}(2004)}]{Feigelson.04}
{Feigelson}, E.~D., \& {Lawson}, W.~A. 2004, The Astrophysical Journal, 614,
  267

\bibitem[{Fischer \& Marcy(1992)}]{Marcy92}
Fischer, D., \& Marcy, G. 1992, ApJ, 396, 178

\bibitem[{{Fitzpatrick}(1993)}]{Fitzpatrick.93}
{Fitzpatrick}, M.~J. 1993, in Astronomical Society of the Pacific Conference
  Series, Vol.~52, Astronomical Data Analysis Software and Systems II, ed.
  R.~J. {Hanisch}, R.~J.~V. {Brissenden}, \& J.~{Barnes}, 472

\bibitem[{{Follette} {et~al.}(2013){Follette}, {Tamura}, {Hashimoto},
  {Whitney}, {Grady}, {Close}, {Andrews}, {Kwon}, {Wisniewski}, {Brandt},
  {Mayama}, {Kandori}, {Dong}, {Abe}, {Brandner}, {Carson}, {Currie}, {Egner},
  {Feldt}, {Goto}, {Guyon}, {Hayano}, {Hayashi}, {Hayashi}, {Henning},
  {Hodapp}, {Ishii}, {Iye}, {Janson}, {Knapp}, {Kudo}, {Kusakabe}, {Kuzuhara},
  {McElwain}, {Matsuo}, {Miyama}, {Morino}, {Moro-Martin}, {Nishimura}, {Pyo},
  {Serabyn}, {Suto}, {Suzuki}, {Takami}, {Takato}, {Terada}, {Thalmann},
  {Tomono}, {Turner}, {Watanabe}, {Yamada}, {Takami}, \& {Usuda}}]{Follette.13}
{Follette}, K.~B., {et~al.} 2013, ApJ, 767, 10

\bibitem[{Forrest {et~al.}(2004)Forrest, Sargent, Furlan, D'Alessio, Calvet,
  Hartmann, Uchida, Green, Watson, Chen, Kemper, Keller, Sloan, Herter, Brandl,
  Houck, Barry, Hall, Morris, Najita, \& Myers}]{Forrest.04}
Forrest, W.~J., {et~al.} 2004, ApJS, 154, 443

\bibitem[{Furlan {et~al.}(2009)Furlan, Watson, McClure, Manoj, Espaillat,
  D'Alessio, Calvet, Kim, Sargent, Forrest, \& Hartmann}]{Furlane.09}
Furlan, E., {et~al.} 2009, ApJ, 703, 1964

\bibitem[{Geers {et~al.}(2007)Geers, van Dishoeck, Visser, Pontoppidan,
  Augereau, Habart, \& Lagrange}]{Geers.07}
Geers, V.~C., van Dishoeck, E.~F., Visser, R., Pontoppidan, K.~M., Augereau,
  J.-C., Habart, E., \& Lagrange, A.~M. 2007, A\&A, 476, 279

\bibitem[{{Ghez} {et~al.}(1997){Ghez}, {McCarthy}, {Patience}, \&
  {Beck}}]{Ghez.97}
{Ghez}, A.~M., {McCarthy}, D.~W., {Patience}, J.~L., \& {Beck}, T.~L. 1997,
  ApJ, 481, 378

\bibitem[{Ghez {et~al.}(1993)Ghez, Neugebauer, \& Matthews}]{Ghez.93}
Ghez, A.~M., Neugebauer, G., \& Matthews, K. 1993, AJ, 106, 2005

\bibitem[{{Gizis}(1997)}]{Gizis.97}
{Gizis}, J.~E. 1997, AJ, 113, 806

\bibitem[{Gorti {et~al.}(2009)Gorti, Dullemond, \& Hollenbach}]{Gorti.09}
Gorti, U., Dullemond, C.~P., \& Hollenbach, D. 2009, ApJ, 705, 1237

\bibitem[{{Gray}(1976)}]{Gray.76}
{Gray}, D.~F. 1976, {The observation and analysis of stellar photospheres}

\bibitem[{{Gray}(2005)}]{Gray.05}
---. 2005, {The Observation and Analysis of Stellar Photospheres}

\bibitem[{{Hamilton} {et~al.}(2012){Hamilton}, {Johns-Krull}, {Mundt},
  {Herbst}, \& {Winn}}]{Hamilton.12}
{Hamilton}, C.~M., {Johns-Krull}, C.~M., {Mundt}, R., {Herbst}, W., \& {Winn},
  J.~N. 2012, ApJ, 751, 147

\bibitem[{Hatzes(2002)}]{Hatzes.02}
Hatzes, A. 2002, Astronomische Nachrichten, 323, 392

\bibitem[{{Hilditch}(2001)}]{Hilditch.01}
{Hilditch}, R.~W. 2001, {An Introduction to Close Binary Stars}

\bibitem[{{Hughes} {et~al.}(2009){Hughes}, {Andrews}, {Espaillat}, {Wilner},
  {Calvet}, {D'Alessio}, {Qi}, {Williams}, \& {Hogerheijde}}]{Hughes.09}
{Hughes}, A.~M., {et~al.} 2009, ApJ, 698, 131

\bibitem[{Hughes {et~al.}(2010)Hughes, Andrews, Wilner, Meyer, Carpenter, Qi,
  Hales, Casassus, Hogerheijde, Mamajek, Wolf, Henning, \&
  Silverstone}]{Hughes.10}
Hughes, A.~M., {et~al.} 2010, AJ, 140, 887

\bibitem[{{Hughes} {et~al.}(2007){Hughes}, {Wilner}, {Calvet}, {D'Alessio},
  {Claussen}, \& {Hogerheijde}}]{Hughes.07}
{Hughes}, A.~M., {Wilner}, D.~J., {Calvet}, N., {D'Alessio}, P., {Claussen},
  M.~J., \& {Hogerheijde}, M.~R. 2007, ApJ, 664, 536

\bibitem[{Ireland \& Kraus(2008)}]{Kraus08}
Ireland, M.~J., \& Kraus, A.~L. 2008, ApJ, 678, L59

\bibitem[{{James} {et~al.}(2006){James}, {Melo}, {Santos}, \&
  {Bouvier}}]{James.06}
{James}, D.~J., {Melo}, C., {Santos}, N.~C., \& {Bouvier}, J. 2006, A\& A, 446,
  971

\bibitem[{Jayawardhana {et~al.}(2003)Jayawardhana, Mohanty, \&
  Basri}]{Jaywardhana.03}
Jayawardhana, R., Mohanty, S., \& Basri, G. 2003, The Astrophysical Journal,
  592, 282

\bibitem[{{Johns-Krull} \& {Valenti}(2001)}]{Johns.01}
{Johns-Krull}, C.~M., \& {Valenti}, J.~A. 2001, ApJ, 561, 1060

\bibitem[{Kelson(2003)}]{Kelson03}
Kelson, D. 2003, PASP, 115, 688

\bibitem[{Kim {et~al.}(2009)Kim, Watson, Manoj, Furlan, Najita, Forrest,
  Sargent, Espaillat, Calvet, Luhman, McClure, Green, \& Harrold}]{Kim.09}
Kim, K.~H., {et~al.} 2009, ApJ, 700, 1017

\bibitem[{Kimeswenger {et~al.}(2004)Kimeswenger, Lederle, Richichi, Percheron,
  Paresce, Armsdorfer, Bacher, Cabrera-Lavers, Kausch, Rassia, Schmeja, Tapken,
  Fouque, Maury, \& Epchtein}]{Kimeswenger.04}
Kimeswenger, S., {et~al.} 2004, A\&A, 413, 1037

\bibitem[{{Kiraga}(2012)}]{Kiraga.12}
{Kiraga}, M. 2012, Acta Astronomica, 62, 67

\bibitem[{Knacke {et~al.}(1973)Knacke, Strom, Strom, Young, \&
  Kunkel}]{Knacke.73}
Knacke, R.~F., Strom, K.~M., Strom, S.~E., Young, E., \& Kunkel, W. 1973, ApJ,
  179, 847

\bibitem[{Kraus {et~al.}(2012)Kraus, Ireland, Hillenbrand, \&
  Martinache}]{Kraus.12}
Kraus, A.~L., Ireland, M.~J., Hillenbrand, L.~A., \& Martinache, F. 2012, ApJ,
  745, 19 (11pp)

\bibitem[{Kurosawa {et~al.}(2006)Kurosawa, Harries, \&
  Littlefair}]{Kurosawa.06}
Kurosawa, R., Harries, T.~J., \& Littlefair, S.~P. 2006, MNRAS, 372, 1879

\bibitem[{Lafreniere {et~al.}(2008)Lafreniere, Jayawardhana, Brandeker, Ahmic,
  \& van Kerkwijk}]{Lafreniere.08}
Lafreniere, D., Jayawardhana, R., Brandeker, A., Ahmic, M., \& van Kerkwijk,
  M.~H. 2008, ApJ, 683, 844

\bibitem[{{Lodieu} {et~al.}(2014){Lodieu}, {P{\'e}rez-Garrido}, {B{\'e}jar},
  {Gauza}, {Ruiz}, {Rebolo}, {Pinfield}, \& {Mart{\'{\i}}n}}]{Lodieu.14}
{Lodieu}, N., {P{\'e}rez-Garrido}, A., {B{\'e}jar}, V.~J.~S., {Gauza}, B.,
  {Ruiz}, M.~T., {Rebolo}, R., {Pinfield}, D.~J., \& {Mart{\'{\i}}n}, E.~L.
  2014, A\&A, 569, A120

\bibitem[{{Lopez Mart{\'{\i}}} {et~al.}(2013){Lopez Mart{\'{\i}}}, {Jimenez
  Esteban}, {Bayo}, {Barrado}, {Solano}, \& {Rodrigo}}]{Lopez.13}
{Lopez Mart{\'{\i}}}, B., {Jimenez Esteban}, F., {Bayo}, A., {Barrado}, D.,
  {Solano}, E., \& {Rodrigo}, C. 2013, A\&A, 551, A46

\bibitem[{M.~Lombardi \& Alves(2008)}]{M.-LombardiAlves08}
M.~Lombardi, C. J.~L., \& Alves, J. 2008, A\&A, 480, 785

\bibitem[{{Magazzu} {et~al.}(1992){Magazzu}, {Rebolo}, \&
  {Pavlenko}}]{Magazzu.92}
{Magazzu}, A., {Rebolo}, R., \& {Pavlenko}, I.~V. 1992, ApJ, 392, 159

\bibitem[{{Mamajek}(2009)}]{Mamajek.09}
{Mamajek}, E.~E. 2009, in American Institute of Physics Conference Series, Vol.
  1158, American Institute of Physics Conference Series, ed. T.~{Usuda},
  M.~{Tamura}, \& M.~{Ishii}, 3--10

\bibitem[{{Marcy} \& {Benitz}(1989)}]{Marcy.89}
{Marcy}, G.~W., \& {Benitz}, K.~J. 1989, \apj, 344, 441

\bibitem[{Martin {et~al.}(1999{\natexlab{a}})Martin, Basri, \&
  Zapatero~Osorio}]{Martin.99}
Martin, E.~L., Basri, G., \& Zapatero~Osorio, M.~R. 1999{\natexlab{a}}, AJ,
  118, 1005

\bibitem[{Martin {et~al.}(1999{\natexlab{b}})Martin, Delfosse, Basri, Goldman,
  Forveille, \& Osorio}]{Martin.99b}
Martin, E.~L., Delfosse, X., Basri, G., Goldman, B., Forveille, T., \& Osorio,
  M. R.~Z. 1999{\natexlab{b}}, AJ, 118, 2466

\bibitem[{Melo(2003)}]{Melo03}
Melo, C. H.~F. 2003, A\&A, 410, 269

\bibitem[{{Mohanty} {et~al.}(2002){Mohanty}, {Basri}, {Shu}, {Allard}, \&
  {Chabrier}}]{Mohanty.02}
{Mohanty}, S., {Basri}, G., {Shu}, F., {Allard}, F., \& {Chabrier}, G. 2002,
  ApJ, 571, 469

\bibitem[{Monet {et~al.}(2003)Monet, Levine, Canzian, Bird, Dahn, Guetter,
  Harris, Henden, Leggett, Levison, Luginbuhl, Martini, Munn, Pier, Rhodes,
  Riepe, Vrba, Walker, Westerhout, Reid, Read, \& Tritton}]{Monet.03}
Monet, D.~G., {et~al.} 2003, ApJ, 125, 984

\bibitem[{{Muzerolle} {et~al.}(2010){Muzerolle}, {Allen}, {Megeath},
  {Hern{\'a}ndez}, \& {Gutermuth}}]{Muzerolle.10}
{Muzerolle}, J., {Allen}, L.~E., {Megeath}, S.~T., {Hern{\'a}ndez}, J., \&
  {Gutermuth}, R.~A. 2010, ApJ, 708, 1107

\bibitem[{Natta {et~al.}(2004)Natta, Testi, Muzerolle, Randich, Comeron, \&
  Persi}]{Natta.04}
Natta, A., Testi, L., Muzerolle, J., Randich, S., Comeron, F., \& Persi, P.
  2004, A\&A, 424, 603

\bibitem[{{Neugebauer} {et~al.}(1984){Neugebauer}, {Habing}, {van Duinen},
  {Aumann}, {Baud}, {Beichman}, {Beintema}, {Boggess}, {Clegg}, {de Jong},
  {Emerson}, {Gautier}, {Gillett}, {Harris}, {Hauser}, {Houck}, {Jennings},
  {Low}, {Marsden}, {Miley}, {Olnon}, {Pottasch}, {Raimond}, {Rowan-Robinson},
  {Soifer}, {Walker}, {Wesselius}, \& {Young}}]{Neugebauer.84}
{Neugebauer}, G., {et~al.} 1984, ApJL, 278, L1

\bibitem[{Neuhauser \& Forbrich(2008)}]{Forbrich08}
Neuhauser, \& Forbrich. 2008, Handbook of Star Forming Regions, Volume II: The
  Southern Sky ASP Monograph Publications, Vol. 5. Edited by Bo Reipurth, 5,
  735

\bibitem[{{Neuh{\"a}user} {et~al.}(2000){Neuh{\"a}user}, {Walter}, {Covino},
  {Alcal{\'a}}, {Wolk}, {Frink}, {Guillout}, {Sterzik}, \&
  {Comer{\'o}n}}]{Neuhauser.00}
{Neuh{\"a}user}, R., {et~al.} 2000, A\&AS, 146, 323

\bibitem[{Nguyen {et~al.}(2009)Nguyen, Scholz, van Kerkwijk, Jayawardhana, \&
  Brandeker}]{Nguyen.09}
Nguyen, D.~C., Scholz, A., van Kerkwijk, M.~H., Jayawardhana, R., \& Brandeker,
  A. 2009, ApJL, 694, L153

\bibitem[{Nidever {et~al.}(2002)Nidever, Marcy, Butler, \& Vogt}]{Nidever.02}
Nidever, D.~L., Marcy, G.~W., Butler, R. Paul;~Fischer, D.~A., \& Vogt, S.~S.
  2002, ApJS, 141, 503

\bibitem[{{Orosz} {et~al.}(2012){Orosz}, {Welsh}, {Carter}, {Fabrycky},
  {Cochran}, {Endl}, {Ford}, {Haghighipour}, {MacQueen}, {Mazeh},
  {Sanchis-Ojeda}, {Short}, {Torres}, {Agol}, {Buchhave}, {Doyle}, {Isaacson},
  {Lissauer}, {Marcy}, {Shporer}, {Windmiller}, {Barclay}, {Boss}, {Clarke},
  {Fortney}, {Geary}, {Holman}, {Huber}, {Jenkins}, {Kinemuchi}, {Kruse},
  {Ragozzine}, {Sasselov}, {Still}, {Tenenbaum}, {Uddin}, {Winn}, {Koch}, \&
  {Borucki}}]{Orosz.12}
{Orosz}, J.~A., {et~al.} 2012, Science, 337, 1511

\bibitem[{{Padgett} {et~al.}(2008){Padgett}, {Rebull}, {Stapelfeldt},
  {Chapman}, {Lai}, {Mundy}, {Evans}, {Brooke}, {Cieza}, {Spiesman},
  {Noriega-Crespo}, {McCabe}, {Allen}, {Blake}, {Harvey}, {Huard},
  {J{\o}rgensen}, {Koerner}, {Myers}, {Sargent}, {Teuben}, {van Dishoeck},
  {Wahhaj}, \& {Young}}]{Padgett.08}
{Padgett}, D.~L., {et~al.} 2008, ApJ, 672, 1013

\bibitem[{{Pinilla} {et~al.}(2012){Pinilla}, {Benisty}, \&
  {Birnstiel}}]{Pinilla.12}
{Pinilla}, P., {Benisty}, M., \& {Birnstiel}, T. 2012, A\&A, 545, A81

\bibitem[{Pott {et~al.}(2010)Pott, Furlan, Herbst, \& Metchev}]{Pott.10}
Pott, Jorg-Uwe;~Perrin, M.~D., Furlan, Elise;~Ghez, A.~M., Herbst, T.~M., \&
  Metchev, S. 2010, ApJ, 710, 265

\bibitem[{Prato(2007)}]{Prato07}
Prato, L. 2007, ApJ, 657

\bibitem[{{Prato} {et~al.}(2003){Prato}, {Greene}, \& {Simon}}]{Prato.03}
{Prato}, L., {Greene}, T.~P., \& {Simon}, M. 2003, ApJ, 584, 853

\bibitem[{{Qian} {et~al.}(2012){Qian}, {Liu}, {Zhu}, {Dai}, {Fern{\'a}ndez
  Laj{\'u}s}, \& {Baume}}]{Qian2.12}
{Qian}, S.-B., {Liu}, L., {Zhu}, L.-Y., {Dai}, Z.-B., {Fern{\'a}ndez
  Laj{\'u}s}, E., \& {Baume}, G.~L. 2012, MNRAS

\bibitem[{Qian {et~al.}(2012b)Qian, Zhu, Dai, Fernández-Lajús, Xiang, \&
  He}]{Qian.12}
Qian, S.-B., Zhu, L.-Y., Dai, Z.-B., Fernández-Lajús, E., Xiang, F.-Y., \&
  He, J.-J. 2012b, ApJL, 745, L23

\bibitem[{Quillen {et~al.}(2004)Quillen, Blackman, \& Frank}]{Quillen.04}
Quillen, A.~C., Blackman, E.~G., \& Frank, Adam;~Varniere, P. 2004, ApJ, 612,
  L137

\bibitem[{{Raghavan} {et~al.}(2010){Raghavan}, {McAlister}, {Henry}, {Latham},
  {Marcy}, {Mason}, {Gies}, {White}, \& {ten Brummelaar}}]{Raghavan.10}
{Raghavan}, D., {et~al.} 2010, ApJS, 190, 1

\bibitem[{{Rajpurohit} {et~al.}(2013){Rajpurohit}, {Reyl{\'e}}, {Allard},
  {Homeier}, {Schultheis}, {Bessell}, \& {Robin}}]{Raj.13}
{Rajpurohit}, A.~S., {Reyl{\'e}}, C., {Allard}, F., {Homeier}, D.,
  {Schultheis}, M., {Bessell}, M.~S., \& {Robin}, A.~C. 2013, A\&A, 556, A15

\bibitem[{Ratzka {et~al.}(2005)Ratzka, Kohler, \& Leinert}]{Ratzka.05}
Ratzka, T., Kohler, R., \& Leinert, C. 2005, A\&A, 437, 611

\bibitem[{Reid \& Cruz(2002)}]{Reid.02}
Reid, I.~N., \& Cruz, K.~L. 2002, AJ, 123, 2806

\bibitem[{{Reiners} {et~al.}(2001){Reiners}, {Schmitt}, \&
  {K{\"u}rster}}]{Reiners.01}
{Reiners}, A., {Schmitt}, J.~H.~M.~M., \& {K{\"u}rster}, M. 2001, A\&A, 376,
  L13

\bibitem[{{Rice} {et~al.}(2006){Rice}, {Armitage}, {Wood}, \&
  {Lodato}}]{Rice.06}
{Rice}, W.~K.~M., {Armitage}, P.~J., {Wood}, K., \& {Lodato}, G. 2006, MNRAS,
  373, 1619

\bibitem[{Richichi \& Percheron(2002)}]{Percheron02}
Richichi, \& Percheron. 2002, A\&A, 386, 492

\bibitem[{{Richichi} {et~al.}(2005){Richichi}, {Percheron}, \&
  {Khristoforova}}]{Richichi.05}
{Richichi}, A., {Percheron}, I., \& {Khristoforova}, M. 2005, A\&A, 431, 773

\bibitem[{{Rosero} {et~al.}(2011){Rosero}, {Prato}, {Wasserman}, \&
  {Rodgers}}]{Rosero.11}
{Rosero}, V., {Prato}, L., {Wasserman}, L.~H., \& {Rodgers}, B. 2011, AJ, 141,
  13

\bibitem[{{Samus'} {et~al.}(2003){Samus'}, {Goranskii}, {Durlevich}, {Zharova},
  {Kazarovets}, {Kireeva}, {Pastukhova}, {Williams}, \& {Hazen}}]{Samus.03}
{Samus'}, N.~N., {et~al.} 2003, Astronomy Letters, 29, 468

\bibitem[{{Schwamb} {et~al.}(2013){Schwamb}, {Orosz}, {Carter}, {Welsh},
  {Fischer}, {Torres}, {Howard}, {Crepp}, {Keel}, {Lintott}, {Kaib}, {Terrell},
  {Gagliano}, {Jek}, {Parrish}, {Smith}, {Lynn}, {Simpson}, {Giguere}, \&
  {Schawinski}}]{Schwamb.13}
{Schwamb}, M.~E., {et~al.} 2013, ApJ, 768, 127

\bibitem[{{Shkolnik} {et~al.}(2009){Shkolnik}, {Liu}, \& {Reid}}]{Shkolnik.09}
{Shkolnik}, E., {Liu}, M.~C., \& {Reid}, I.~N. 2009, ApJ, 699, 649

\bibitem[{Shkolnik {et~al.}(2010)Shkolnik, Hebb, Liu, \&
  Collier~Cameron}]{Shkolnik.10}
Shkolnik, E.~L., Hebb, L., Liu, Michael C.and~Reid, I.~N., \& Collier~Cameron,
  A. 2010, ApJ, 716, 1522

\bibitem[{{Sicilia-Aguilar} {et~al.}(2006){Sicilia-Aguilar}, {Hartmann},
  {Calvet}, {Megeath}, {Muzerolle}, {Allen}, {D'Alessio}, {Mer{\'{\i}}n},
  {Stauffer}, {Young}, \& {Lada}}]{S-A.06}
{Sicilia-Aguilar}, A., {et~al.} 2006, ApJ, 638, 897

\bibitem[{Sicilia-Aguilar {et~al.}(2008)Sicilia-Aguilar, Henning, Juhasz,
  Bouwman, Garmire, \& Garmire}]{S-A.08}
Sicilia-Aguilar, A., Henning, T., Juhasz, A., Bouwman, J., Garmire, G., \&
  Garmire, A. 2008, ApJ, 687, 1145

\bibitem[{Siess {et~al.}(2000)Siess, Dufour, \& Forestini}]{Seiss.00}
Siess, L., Dufour, E., \& Forestini, M. 2000, A\&A, 358, 593

\bibitem[{{Sim{\'o}n-D{\'{\i}}az} \& {Herrero}(2007)}]{Simon-Diaz.07}
{Sim{\'o}n-D{\'{\i}}az}, S., \& {Herrero}, A. 2007, A\&A, 468, 1063

\bibitem[{{Skrutskie} {et~al.}(2006){Skrutskie}, {Cutri}, {Stiening},
  {Weinberg}, {Schneider}, {Carpenter}, {Beichman}, {Capps}, {Chester},
  {Elias}, {Huchra}, {Liebert}, {Lonsdale}, {Monet}, {Price}, {Seitzer},
  {Jarrett}, {Kirkpatrick}, {Gizis}, {Howard}, {Evans}, {Fowler}, {Fullmer},
  {Hurt}, {Light}, {Kopan}, {Marsh}, {McCallon}, {Tam}, {Van Dyk}, \&
  {Wheelock}}]{Skrutskie.06}
{Skrutskie}, M.~F., {et~al.} 2006, AJ, 131, 1163

\bibitem[{Struve \& Rudkjobing(1949)}]{Rudkjobing49}
Struve, \& Rudkjobing. 1949, ApJ, 109, 92

\bibitem[{Suarez {et~al.}(2006)Suarez, Garcia-Lario, Manchado, Manteiga, Ulla,
  \& Pottasch}]{Suarez.06}
Suarez, O., Garcia-Lario, P., Manchado, A., Manteiga, M., Ulla, A., \&
  Pottasch, S.~R. 2006, A\&A, 458, 173

\bibitem[{{Torres} {et~al.}(2006){Torres}, {Quast}, {da Silva}, {de La Reza},
  {Melo}, \& {Sterzik}}]{Torres.06}
{Torres}, C.~A.~O., {Quast}, G.~R., {da Silva}, L., {de La Reza}, R., {Melo},
  C.~H.~F., \& {Sterzik}, M. 2006, A\&A, 460, 695

\bibitem[{{Vogt} {et~al.}(2012){Vogt}, {Schmidt}, {Neuh{\"a}user}, {Bedalov},
  {Roell}, {Seifahrt}, \& {Mugrauer}}]{Vogt.12}
{Vogt}, N., {Schmidt}, T.~O.~B., {Neuh{\"a}user}, R., {Bedalov}, A., {Roell},
  T., {Seifahrt}, A., \& {Mugrauer}, M. 2012, A\&A, 546, A63

\bibitem[{{Vrba} {et~al.}(1993){Vrba}, {Coyne}, \& {Tapia}}]{Vrba.93}
{Vrba}, F.~J., {Coyne}, G.~V., \& {Tapia}, S. 1993, AJ, 105, 1010

\bibitem[{Wahhaj {et~al.}(2010)Wahhaj, Cieza, Koerner, Stapelfeldt, Padgett,
  Case, Keller, Merin, Evans, Harvey, Sargent, van Dishoeck, Allen, Blake,
  Brooke, Chapman, Mundy, \& Myers}]{Wahhaj.10}
Wahhaj, Z., {et~al.} 2010, ApJ, 724, 835

\bibitem[{White \& Basri(2003)}]{WhiteBasri.03}
White, R.~J., \& Basri, G. 2003, The Astrophysical Journal, 582, 1109

\bibitem[{White {et~al.}(2007)White, Gabor, \& Hillenbrand}]{White07}
White, R.~J., Gabor, J.~M., \& Hillenbrand, L.~A. 2007, AJ, 133, 2524

\bibitem[{{White} {et~al.}(2007){White}, {Gabor}, \& {Hillenbrand}}]{White.07}
{White}, R.~J., {Gabor}, J.~M., \& {Hillenbrand}, L.~A. 2007, AJ, 133, 2524

\bibitem[{{White} \& {Ghez}(2001)}]{White.01}
{White}, R.~J., \& {Ghez}, A.~M. 2001, ApJ, 556, 265

\bibitem[{{White} \& {Hillenbrand}(2005)}]{White.05}
{White}, R.~J., \& {Hillenbrand}, L.~A. 2005, ApJL, 621, L65

\bibitem[{Wilking {et~al.}(2005)Wilking, Meyer, Robinson, \&
  Greene}]{Wilking.05}
Wilking, B.~A., Meyer, M.~R., Robinson, J.~G., \& Greene, T.~P. 2005, AJ, 130,
  1733

\bibitem[{{Wyatt}(1985)}]{Wyatt.85}
{Wyatt}, W.~F. 1985, in Stellar Radial Velocities, ed. A.~G.~D. {Philip} \&
  D.~W. {Latham}, 123--130

\bibitem[{{Zhu} {et~al.}(2011){Zhu}, {Nelson}, {Hartmann}, {Espaillat}, \&
  {Calvet}}]{Zhu.11}
{Zhu}, Z., {Nelson}, R.~P., {Hartmann}, L., {Espaillat}, C., \& {Calvet}, N.
  2011, ApJ, 729, 47

\bibitem[{{Zhu} \& {Stone}(2014)}]{Zhu.12}
{Zhu}, Z., \& {Stone}, J.~M. 2014, ApJ, 795, 53

\bibitem[{{Zickgraf} {et~al.}(2005){Zickgraf}, {Krautter}, {Reffert},
  {Alcal{\'a}}, {Mujica}, {Covino}, \& {Sterzik}}]{Zickgraf.05}
{Zickgraf}, F.-J., {Krautter}, J., {Reffert}, S., {Alcal{\'a}}, J.~M.,
  {Mujica}, R., {Covino}, E., \& {Sterzik}, M.~F. 2005, A\&A, 433, 151

\bibitem[{{Zuckerman} \& {Song}(2004)}]{Zuckerman.04}
{Zuckerman}, B., \& {Song}, I. 2004, ARA\&A, 42, 685

\end{thebibliography}

\end{document}